# Nitrogen-Vacancy Magnetic Relaxometry of Nanoclustered Cytochrome C Proteins

Suvechhya Lamichhane,[1] Rupak Timalsina,[2] Cody Schultz,[3] Ilja Fescenko,[4] Kapildeb Ambal,[5] Sy-Hwang Liou,[1] Rebecca Y. Lai,[3] and Abdelghani Laraoui[1,2,*]

[1]*Department of Physics and Astronomy and the Nebraska Center for Materials and Nanoscience, University of Nebraska-Lincoln, Lincoln, Nebraska 68588, USA*
[2]*Department of Mechanical & Materials Engineering, University of Nebraska-Lincoln, Lincoln, NE 68588, USA*
[3]*Department of Chemistry, University of Nebraska-Lincoln, Lincoln, NE 68588, USA*
[4]*Laser Center, University of Latvia, Riga, LV-1004, Latvia*
[5]*Department of Mathematics, Statistics, and Physics, Wichita State University, Wichita, KS 67260, USA*
[*]*Corresponding Author E-mail: alaraoui2@unl.edu*

**ABSTRACT:** Nitrogen-vacancy (NV) magnetometry offers an alternative tool to detect paramagnetic centers in cells with a favorable combination of magnetic sensitivity and spatial resolution. Here, we employ NV magnetic relaxometry to detect cytochrome C (Cyt-C) nanoclusters. Cyt-C is a water-soluble protein that plays a vital role in the electron transport chain of mitochondria. Under ambient conditions, the heme group in Cyt-C remains in the $Fe^{3+}$ state, which is paramagnetic. We vary the concentration of Cyt-C from 6 to 54 μM and observe a reduction of the NV spin–lattice relaxation time ($T_1$) from 1.2 ms to 150 μs, which is attributed to the spin noise originating from the $Fe^{3+}$ spins. NV $T_1$ imaging of Cyt-C drop-casted on a nanostructured diamond chip allows us to detect the relaxation rates from the adsorbed $Fe^{3+}$ within Cyt-C.

**KEYWORDS:** *Nitrogen-vacancy, Relaxometry, Cytochrome C, Iron, Paramagnetic*

Cytochromes, including cytochrome C (Cyt-C), are redox-active proteins involved in the electron transport chain (ETC) and redox catalysis.[1] It is a water-soluble protein with a molecular weight of 13 kDa typically located in the inter-mitochondrial membrane.[1,2] The main function of Cyt-C within the mitochondria is to serve as an intermediate electron receptor during respiration, facilitating the transfer of electrons between Complex III and Complex IV in the ETC.[1] The electron transfer involves the reduction of the oxidized Cyt-C ($Fe^{+3}$) heme group to $Fe^{+2}$ by an electron from Complex III. Subsequently, Cyt-C releases the electron to Complex IV, and the Fe center returns to the oxidized $Fe^{+3}$ state.[3] During cell necrosis and apoptosis, Cyt-C is released from the mitochondria, which leads to the non-inflammatory activation of apoptotic protease activating factor 1, eventually cascading to caspase-9 activation and the execution-phase of the cell.[4] The concentration of Cyt-C decreases in the mitochondria and increases in the cytosol during apoptosis, resulting in variations of Cyt-C amounts depending on the specific cellular region being analyzed.[1,2]

While the intracellular release of Cyt-C is not associated with inflammation, its release into the extracellular matrix has been correlated with inflammatory conditions and cellular damage. Studies on rats resuscitated after ventricular fibrillation showed a tenfold increase in plasma Cyt-C, with the survival outcome inversely correlated to plasma Cyt-C concentrations.[5] Increased Cyt-C levels have also been linked to liver disease, liver damage, and kidney damage.[6,7] Therefore, plasma Cyt-C can be a useful diagnostic tool for pathogen-associated or danger-associated molecular patterns.[2]

Numerous techniques have been utilized to detect Cyt-C, including electron paramagnetic resonance (EPR) spectrometry[8–10] and X-ray photoelectron spectroscopy (XPS).[11,12] The variation of the Cyt-C concentration in the cells depending on its current physiology state makes it hard to use bulk techniques such as EPR to detect it at the sub-cellular level. Among the currently available high-sensitivity and spatial resolution detection techniques, magnetic imaging based on the nitrogen-vacancy (NV) center in diamond[13–18] emerges as one of the most promising tools for high-resolution nanoscale imaging. The NV center is a spin-1 defect with optically addressable electron spin properties[19–21] and millisecond spin relaxation times ($T_1$, $T_2$) at room temperature.[22] These capabilities have opened new opportunities in quantum sensing,[23] nanoscale magnetometry,[24–26] and biosensing.[27] For example, it enabled the initial detection of single proteins[28] and the realization of nanoscale nuclear magnetic resonance (NMR) spectroscopy.[29–32] NV magnetometry has been used very recently to measure the magnetic properties of individual (size < 1 μm) malarial hemozoin biocrystals[18] and $[Fe(Htrz)_2(trz)](BF_4)$ spin crossover molecules.[13] Another approach of NV magnetometry is to detect spin noise from metal ions through the decrease of the spin-lattice $T_1$ relaxation time. NV $T_1$ relaxometry has been used to detect $Gd^{+3}$ ions,[33] $Cu^{+2}$ ions,[34] and more recently to map $Fe^{+3}$ in ferritin proteins.[16,17] There have been several studies using NV relaxometry in the biomedical field to study free radicals present in single mitochondria,[35] endothelial cells,[36] dendritic cells,[37] and virus-infected cells,[38] providing great insights on understanding the cellular metabolism.

In this study, we employ NV $T_1$ relaxometry to detect the concentration variations in Cyt-C nanoclusters (diameter: 50 – 200 nm, height: 5 – 15 nm) through the paramagnetic $Fe^{+3}$ centers. The changes in Cyt-C concentration directly impact $T_1$, facilitating a quantitative detection of Cyt-C levels. To achieve this, we systematically altered the amount of Cyt-C adsorbed on the diamond chip, ranging from 6 μM to 54 μM through drop-casting. The observed reduction in $T_1$ from ~ 1.2 ms to < 200 μs is attributed to the spin-noise generated by the $Fe^{+3}$ centers present within Cyt-C. NV $T_1$ imaging of Cyt-C nanoclusters, drop-casted on a functionalized nanostructured diamond chip, allows us to deduce the relaxation rates caused by the adsorbed $Fe^{+3}$ of a density in the range of $1.44 \times 10^6$ to $1.7 \times 10^7$ per μm$^2$ within Cyt-C.

The $Fe^{+3}$ center in Cyt-C is coordinated to four nitrogen atoms,[1] forming the heme ring as depicted in Figure 1a. Additionally, $Fe^{+3}$ coordinates with methionine and histidine, resulting in a hexacoordinated configuration.[3] Notably, the sulfur atom from methionine acts as a strong-field ligand, causing the heme complex to exhibit a low spin state ($S = ½$) under ambient conditions.[39] In this work, we use horse heart Cyt-C powder purchased from Sigma Aldrich (molecular weight = 12,384 g/mol). As received, Cyt-C powder is predominantly in the oxidized state $Fe^{+3}$ and diluted in deionized (DI) water with a density of 200 mg/ml (~ 16.1 mM, pH = 5). Cyt-C in its native state remains neither unfolded or denatured within a pH range of 3 to 7.[40,41]

To measure the diameter and height of Cyt-C nanoclusters, we further diluted the Cyt-C solution to a concentration of ~ 0.1 mM and drop-casted 5 μL on silicon and diamond substrates, allowing it to evaporate and form a thin film. To alleviate the charging effects in SEM measurements,[42] we drop-casted Cyt-C nanoclusters on carbon tape[13] and on gold (Au) coated diamond substrates, see the Supporting Information (SI) Section S1. Figure 1b shows the SEM image of nanoclustered Cyt-C proteins on carbon tape with an aggregated diameter that varies from 30 nm to 150 nm (SI Section S1). The diameter of a single monomer of Cyt-C is reported to be ~ 3.4 nm.[43] However, due to surface interactions, when Cyt-C is drop-casted on the diamond surfaces, it tends to agglomerate, resulting in larger nanoclusters. Remarkably, even at a lower concentration (≤ 2 μM), this agglomeration phenomenon is still observed on the diamond substrate with an average diameter of Cyt-C nanoclusters observed 100 ± 50 nm (Figure S1). Similar agglomeration effects have been reported across different concentrations of Cyt-C.[43] Despite its stability, even in its natural state Cyt-C tends to polymerize rather than remain in a monomeric state, possibly through aggregation.[44] AFM measurements of 2 μM Cyt-C solution drop-casted on top of the diamond substrate show a height of ~ 9 ± 6 nm (Figure 1c).

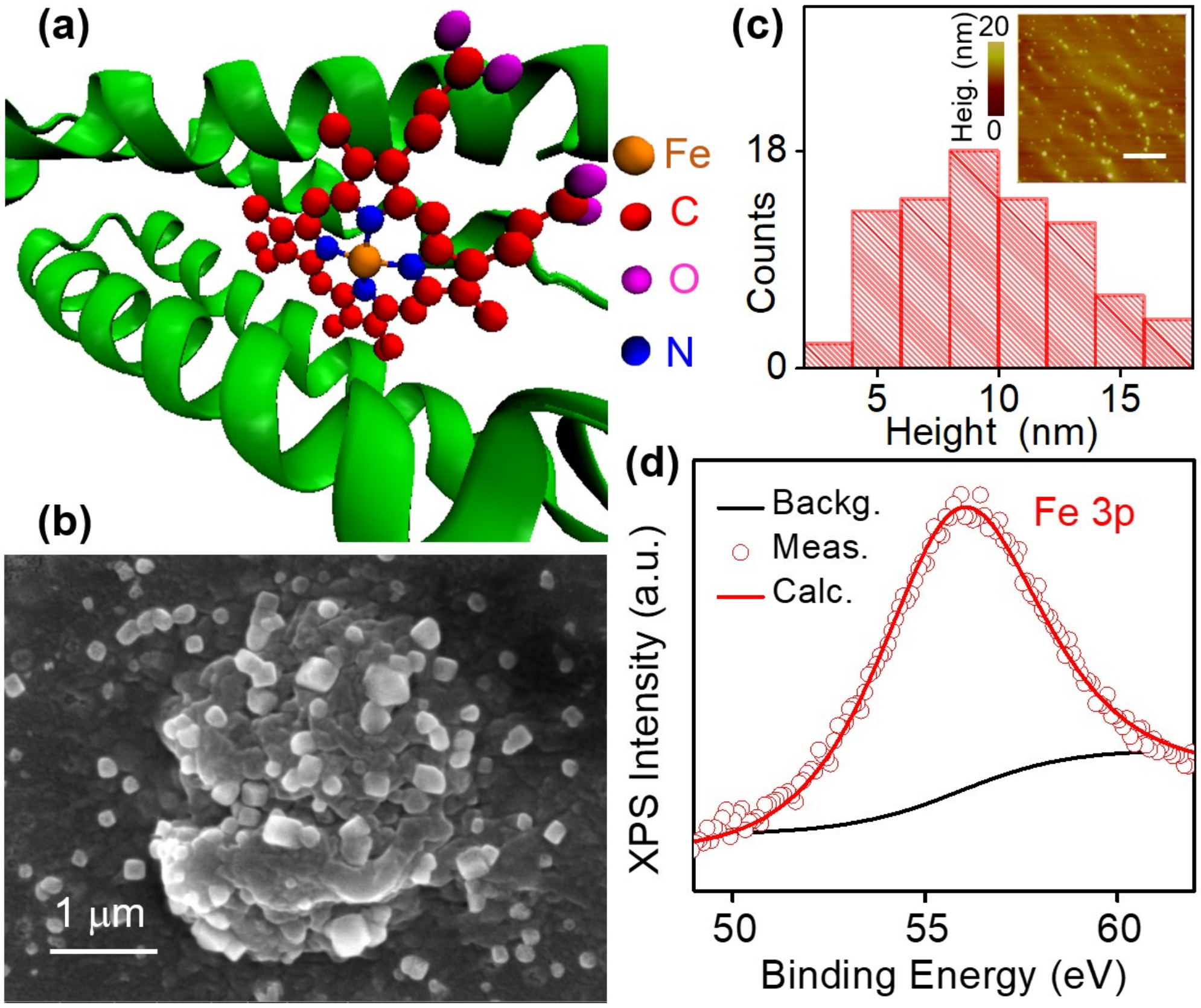

**Figure 1.** (a) Molecular structure of Cyt-C complex derived from reference [1]. (b) SEM image of 18 μM Cyt-C drop-casted on a carbon tape to prevent charging effect. (c) AFM height distribution of Cyt-C nanoclusters (concentration of 2 μM) drop-cased on top of the diamond substrate. The mean height of Cyt-C nanoclusters is ~ 9 ± 6 nm. Inset of (c): AFM image of the Cyt-C nanoclusters (the scale bar is 5 μm). (d) Measured (open circles) XPS spectra of Fe 3p in Cyt-C with the fitting (red solid line) showing the $Fe^{+3}$ state. The background signal is plotted in the black solid line curve.

To further ensure the absence of impurities, XPS measurements were conducted (SI Section S2). We employed a high-resolution XPS by using a step size of 0.1 eV for the analysis of N, S, C, O, and Fe. Following the data acquisition, we conducted peak fitting using a combined Lorentzian and Gaussian function which allowed us to deconvolute the spectral feature and extract information regarding the chemical states and binding energy of the elements. Figure 1d shows the measured (open circles) and calculated (solid line) high-resolution spectra of Fe 3p. The spectra exhibit a distinct peak at a binding energy of 55.6 eV, indicative of the Fe oxidation state being +3.[45] Moreover, we corroborated the presence of Fe-sulfur bonds at binding energy of 167 eV through the examination of the high-resolution sulfur peak (Figure S2d).[46]

The negatively charged NV center in the diamond lattice is a substitutional nitrogen adjacent to a vacancy site (Figure 2b) with a spin triplet in the ground state that features a zero-field splitting $D$ = 2.87 GHz between states $m_s$ = 0 and $m_s$ = ±1 (Figure 2c).[19,21] A green laser illumination (532 nm) yields spin-conserving excitation to the excited triplet state, which in turn leads to far-red photoluminescence (650 – 800 nm). Intersystem crossing to metastable singlet states takes place preferentially for NV centers in the $m_s$ = ±1 states, ultimately resulting in an almost complete transfer of population to the $m_s$ = 0 state.[19] Microwave (MW) excitation allows spin transitions from $m_s$ = 0 to $m_s$ = ±1 sublevels. The applied magnetic field breaks the degeneracy of the $m_s$ = ±1, leading to a pair of spin transitions ($m_s$ = 0 to $m_s$ = +1 and $m_s$ = 0 to $m_s$ = -1) that can be interrogated *via* optically detected magnetic resonance (ODMR) spectroscopy.[13,18] This method results in a set of resonance peaks that are sensitive to magnetic field (Figure S5b). In addition, if the laser excitation is abruptly turned off, the amplitude of the ODMR signal decays exponentially as the NV electron spins relax from their polarized states. The rate of this decay ($\Gamma_1$) or relaxation time ($T_1 = 1/\Gamma_1$) depends on the weak random magnetic fields created by the diamond sensor itself and the external spin noise.[47,48] Measurements of the rate change in $\Gamma_1$ is the basis of NV-$T_1$ relaxometry which is used here to detect magnetic dipole-dipole interaction between NV spins and the fluctuating $Fe^{+3}$ spins in Cyt-C nanoclusters.[49]

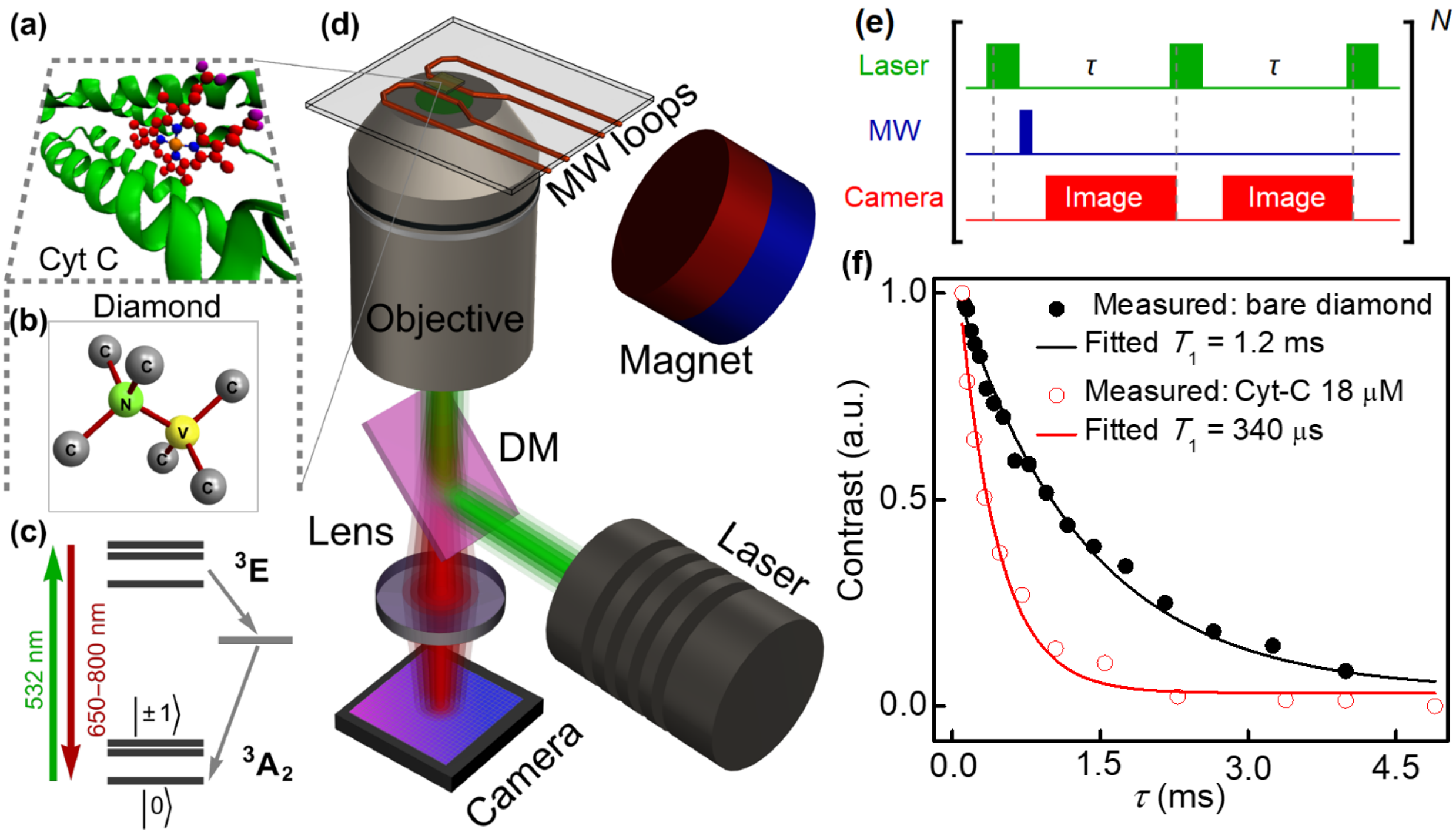

**Figure 2.** (a) Molecular structure of Cyt-C protein. (b) A schematic of the NV center inside the diamond lattice (nitrogen: green atom, yellow: vacancy). (c) A schematic of the energy levels of the NV center ground ($^3A^2$) and excited ($^3E$) states with intermediate metastable state. A green laser (532 nm) initializes the NV center spins and results in fluorescence in wavelength range of 650 – 800 nm. (d) A schematic of the widefield NV microscope used for $T_1$ relaxometry measurements and imaging of Cyt-C nanoclusters. DM is the dichroic mirror. (e) A schematic of a pulse sequence for $T_1$ relaxometry imaging. (f) $T_1$ curve measured on a bare diamond (filled circles) and after drop-casting 18 μM Cyt-C solution on diamond (open circles). The black and red solid lines are exponential decay function fits of the measured $T_1$ relaxation curves.

To make the NV sensor, we used 3 mm × 2.5 mm × 0.05 mm electronic grade (100) cut and polished diamond. A thin (~ 5 – 8 nm) NV sensing layer (Figure S5a) was created near the diamond surface by using $^{15}N^{+}$ implantation followed by high vacuum and high temperature annealing and then cleaning in a boiling tri-acid mixture (see SI Section S5 for more details).[18,31,50] To attach Cyt-C nanoclusters, we carboxylated the diamond surface with NV layer underneath (see SI Section S3). The adsorption of Cyt-C occurs through electrostatic attraction between the anionic groups ($-COO^{-}$) present in the carboxylated surface and the positively charged amino groups ($-NH^{3+}$), or through the formation of hydrogen bonds between $-NH^{3+}$ and $CO^{-}$.[51] Various concentrations of Cyt-C in DI water were drop-casted on the diamond substrate. We used DI water instead of other buffers to prevent any effects on $T_1$ from the salts and other compounds present in them.

In Figure 2d, we show the schematic of the widefield ODMR microscope. We used a 532 nm laser (power = 180 mW) to excite the NVs over an area of ~ 36 μm × 36 μm, and the NV fluorescence (650 – 800 nm) is mapped onto a sCMOS camera.[13,18] The diamond chip was placed on top of a glass coverslip with patterned gold loops for MW excitation (Figure 2d). A magnetic field $B_{app}$ (3.2 mT) is applied along [111] direction of the (100) diamond enabling the separation of the ODMR peaks for $m_s = 0$ to $m_s = -1$ and $m_s = 0$ to $m_s = +1$ spin transitions (Figure S5b). Detailed information regarding the experimental setup is provided in SI Section S5.

The $T_1$-realxometry measurement protocol is depicted in Figure 2e and summarized here:[47,52] A laser pulse (5 μs) is used first to initialize the NV spins in the $m_s = 0$ state, then a MW $\pi$ pulse is applied to flip the NV spins to $m_s = -1$ state, and finally, a readout laser pulse (5 μs) is applied after varying the measurement time $\tau$. This sequence is followed by another sequence but without $\pi$ pulse. The method and underlying physics are explained in references [47] and [52]. To optimize the timing, each readout laser pulse is used as an initialization pulse for the next sequence. The NV fluorescence is detected by a sCMOS camera, where exposure (4 ms) starts well before the laser pulse, so that the camera stops reading after the first microsecond of the laser induced fluorescence pulse. The first microsecond contains the ODMR contrast signal, which is maximal when the spins are just initialized into $m_s = -1$ or $m_s = 0$ states but decrease with increasing $\tau$ due to the relaxation of the NV spin polarization. We extract this contrast by pixelwise subtracting frames for NVs initialized into $m_s = -1$ (with $\pi$ pulse) from NVs initialized into $m_s = 0$ state (without $\pi$ pulse) then dividing them pixelwise by a sum of both frames. The pair of measurements is repeated $N = 10^4$ times for each value of the varying time $\tau$ (Figures S6b, S6c, and S6d). Figure 2f shows preliminary measurements of the ODMR contrast *vs* $\tau$ on bare diamond (filled circles) and after drop-casting 18 μM Cyt-C solution on diamond (open circles). The NV measurements are integrated over an area of 1296 μm$^2$, fitted with one exponential decay function to extract $T_1$ values (see SI Section S6). A reduction of $T_1$ from ~ 1.2 ms for the bare diamond to 340 μs after adding 18 μM Cyt-C is obtained and attributed to the spin-noise originating from $Fe^{+3}$ spins present within the Cyt-C proteins.

To differentiate the effect of $Fe^{+3}$ spins within the Cyt-C nanoclusters from other spin noises due to paramagnetic impurities[53,54] inside the diamond lattice and the diamond surface noise,[55] we performed $T_1$ relaxometry imaging on a nanostructured diamond chip with SiN grating (a separation distance of 4 μm and height of 50 nm). Such a grating is very robust against acids/chemicals during the repeated cleaning process in contrast to other polymers (*e.g.*, polymethyl methacrylate) which would degrade quickly.[34] See the SI Section S4 for the fabrication details. Regions covered with SiN act as barriers for direct dipolar interaction between NV centers and any spin noise on the diamond surface. We found that the SiN film reduces the $\Gamma_1$ rate of NV spins underneath (Figures 3b and 3f) in comparison to conductive metal films (*e.g.* Al) where a strong $\Gamma_1$ increase was observed.[56]

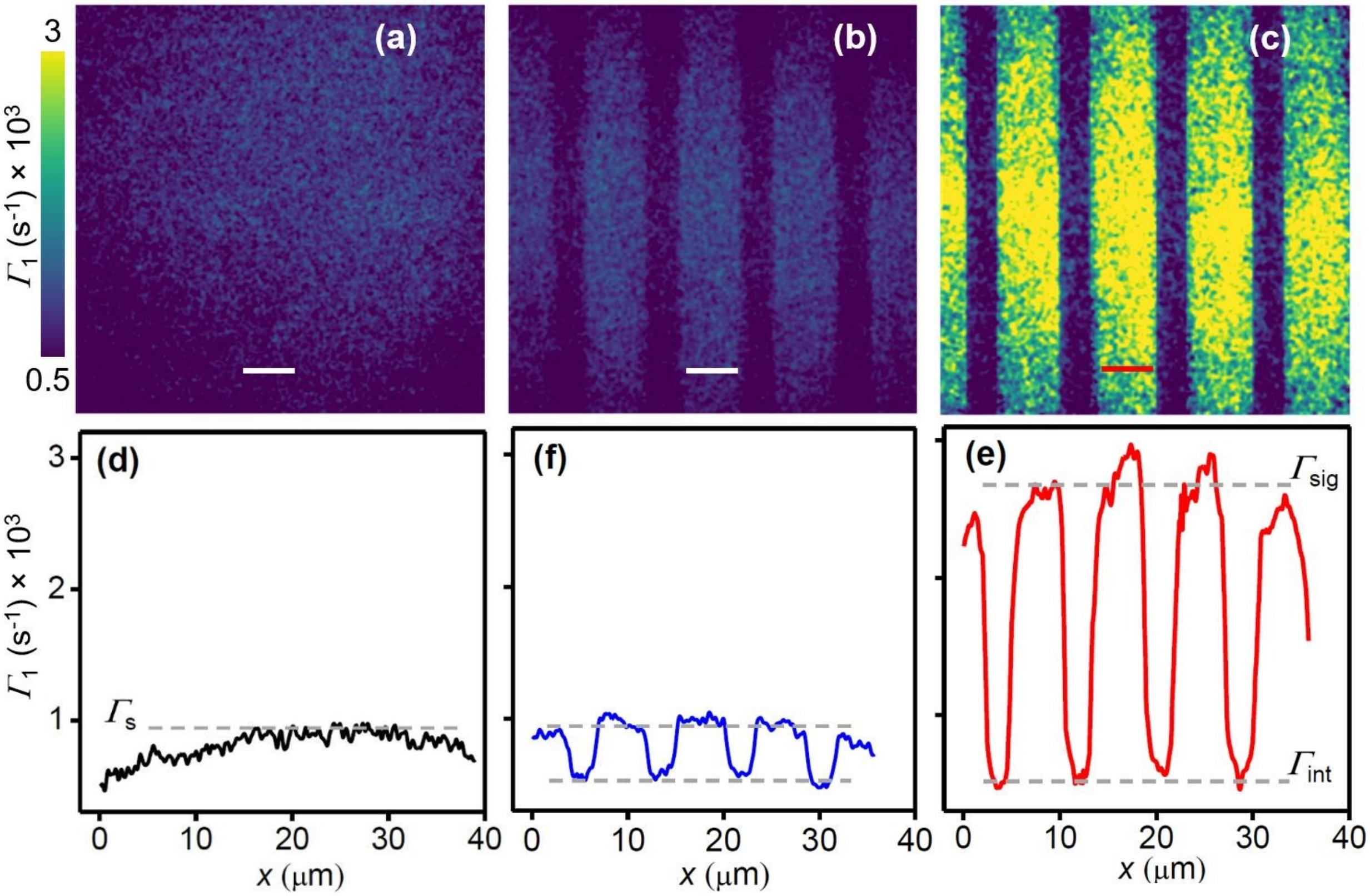

**Figure 3.** $\Gamma_1$ maps acquired by pixelwise exponential fitting of series of maps of the ODMR contrast decays for (a) clean bare diamond, (b) clean diamond with the SiN grating, and (c) Cyt-C with a concentration of 54 μM drop-casted on the diamond with SiN grating. The scale bar in (a), (b), and (c) is 5 μm. Corresponding extracted horizontal $\Gamma_1$ profiles from the measurements done on bare diamond (d), diamond with SiN grating and no Cyt-C (f), and diamond with SiN and Cyt-C (e), respectively.

Figures 3a, 3b, and 3c show $\Gamma_1$ images of the bare diamond, diamond with SiN grating (no Cyt-C), and diamond with SiN and Cyt-C, respectively. The $\Gamma_1$ images were obtained from a series of contrast decay images by pixel-by-pixel fitting with a single exponential function. For further analysis, we integrated $\Gamma_1$ images vertically to measure the $\Gamma_1$ profiles of the measured signals on bare diamond (Figure 3d), diamond with SiN grating and no Cyt-C (Figure 3f), and diamond with SiN and Cyt-C (Figure 3e). The maximum value of $\Gamma_1 = \Gamma_s$ (Figures 3d and 3f) obtained from uncovered clean areas of the NV sensors is $(0.91 \pm 0.05) \times 10^3$ s$^{-1}$. This value decreased towards the edges of the measurement spot due to a slight drop in the intensity of the laser beam, leading to a decrease in the ODMR contrast, which changed the estimation of $\Gamma_1$. To avoid this effect, we estimated the relaxation rates in the central part of the measurement fluorescence spot, where the laser intensity is relatively homogenous. The relaxation rate $\Gamma_{int}$ is $(0.45 \pm 0.05) \times 10^3$ s$^{-1}$ under the SiN gratings, which significantly cancels the surface spin noise of the NV sensor. The relaxation rate $\Gamma_{sig}$ in the presence of Cyt-C nanoclusters with a concentration of 54 μM is $(2.8 \pm 0.2) \times 10^3$ s$^{-1}$ (Figure 3e). Note that the relaxation rate $\Gamma_s$ of our specific diamond sensor with very shallow NV centers is the sum of the intrinsic relaxation ($\Gamma_{int}$) and the relaxation due to surface spin noise ($\Gamma_{sur}$), which accounted for almost half of the relaxation of $\Gamma_s$. However, the NV

relaxation rate in the presence of Cyt-C is $\Gamma_{sig} = \Gamma_{int} + \Gamma_{sur} + \Gamma_{ext}$. From the measurements with very low concentrations of Cyt-C proteins (Figure 4d), we found that $\Gamma_{sur}$ is significantly suppressed by Cyt-C and could be neglected, so that $\Gamma_{ext} \approx \Gamma_{sig} - \Gamma_{int}$ .

Cyt-C, being a heme protein with a $Fe^{+3}$ center, can produce fluctuating magnetic fields that interact with NV spins via dipolar-magnetic-interactions.[33,34,49,57] Figures 4a, 4b, and 4c show the $\Gamma_1$ images of Cyt-C solution drop-casted onto the diamond substrate with a concentration of 25 μM, 18 μM, and 11 μM, respectively. Figure 4d displays the measured relaxation rate $\Gamma_{ext}$ (open circles) as a function of the number of $Fe^{+3}$ centers within Cyt-C adsorbed on the diamond surface over 1 $\mu m^2$ area. To calibrate our measurements and deduce the NV standoff *d* from the diamond surface, we measured the relaxation rate $\Gamma_1$ in $CuSO_4$ solutions of different concentrations on the diamond substrate and found *d* of 5.5 nm (see Figure S8 and SI section S8). After that, we calculated the theoretical dependence of $\Gamma_{ext}$ *versus* the density of $Fe^{+3}$ adsorbed centers per 1 $\mu m^2$ (solid line in Figure 4d) by using Equation S1 while keeping the mean dipolar magnetic coupling strength between NV and $Fe^{+3}$ spins in Cyt-C, $\langle B^2 \rangle$, as a free parameter (SI Section S8). For a spin density of $1.7 \times 10^7$ $Fe^{+3}$ adsorbed/ $\mu m^2$ we determined $\langle B^2 \rangle$ of 0.084 mT.

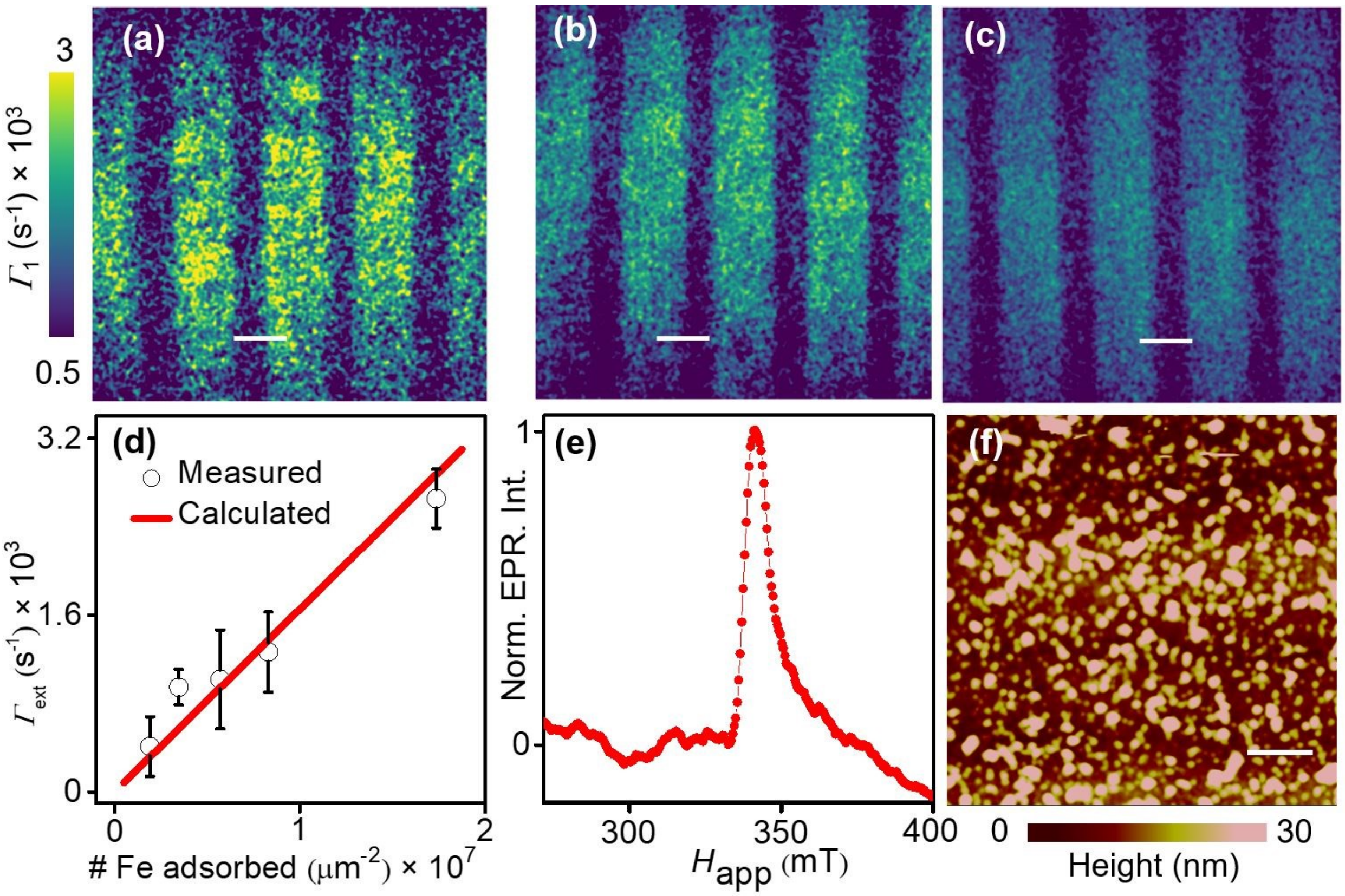

**Figure 4.** (a) NV $\Gamma_1$ map of Cyt-C nanoclusters drop-casted on diamond with a concentration of 25 μM. The bright regions marked the increased relaxation rate due in presence of Cyt-C in contact (no SiN grating) with diamond. The dark regions have SiN due to which the relaxation rate of NV spins gets unaffected. The scale bar is 5 μm. (b) and (c) are the $\Gamma_1$ maps in the presence of 18 μM and 11 μM drop-casted Cyt-C nanoclusters, respectively. (d) Measured (open circles) and calculated (solid line) relaxation rate $\Gamma_1 = \Gamma_{ext}$ of NV spins as function of the density of Fe spins adsorbed on the diamond substrate over 1 $\mu m^2$ area. (e) Measured EPR spectrum of pelletized 15 mg Cyt-C powder. (f) AFM image (30 μm × 30 μm) taken on diamond with Cyt-C nanoclusters with a concentration of 54 μM. The AFM image shows the uniform adsorption of Cyt-C proteins on the diamond surface.

EPR spectroscopy was employed to determine the *g*-factor of $Fe^{+3}$ present in Cyt-C proteins and $Fe^{+3}$ transverse spin relaxation $T_2^*$ rate ($\Gamma_2^* = 1/T_2^*$) by analyzing the linewidth of the EPR spectrum (Figure S7). To improve the signal-to-noise ratio of the EPR signal, we generated pellets from the Cyt-C powder by compressing it into a volume of 20 $mm^3$. We determined an EPR full width at half maximum (FWHM) of 71.43 mT for the pelletized 15 mg Cyt-C powder, corresponding to $\Gamma_2^*$ ~ 2 GHz of the $Fe^{+3}$ spins.[34] We also used AFM imaging as an additional method to estimate the $Fe^{+3}$ spin density within Cyt-C, similar to the approach used in reference [49]. Figure 4f shows the AFM image of 30 μm × 30 μm area of Cyt-C nanoclusters at a concentration of 54 μM, drop-casted on the functionalized diamond surface (see Figure 3c for the corresponding $\Gamma_1$ image). By measuring the diameter and height of Cyt-C nanoclusters from the AFM image we obtained the volume and therefore the concentration of the drop-casted Cyt-C proteins. For instance, 4 μL of 54 μM Cyt-C was drop-casted on the diamond surface (Figure 4f) for which we obtained a spin density of ~ $1.7 \times 10^7$ adsorbed Fe/$\mu m^2$. This value agrees well with the calculated $Fe^{+3}$ spin density value based on the NV imaged area of Cyt-C nanoclusters (36 μm × 36 μm).

In conclusion, we demonstrated the variation in concentration of Cyt-C nanocluster (diameter: 50 – 200 nm, height: 5 – 15 nm) using NV $T_1$ relaxometry through the spin-noise generated by $Fe^{+3}$ spins present in Cyt-C proteins. By patterning the diamond with SiN grating, we performed $T_1$ relaxometry imaging of Cyt-C nanoclusters with different concentrations. The measured NV relaxation rates agree well with the calculated values deduced from a model of interacting $Fe^{+3}$ paramagnetic centers with NV spins in the diamond substrate with a standoff *d* of 5.5 nm. These results open the door to detect Cyt-C on surface-immobilized microbial cells, with the goal of elucidating the effects of Fe, both in ferrous ($Fe^{+2}$) and ferric ($Fe^{+3}$) states, on the electron transport processes during metabolism. The integration of microfluidics[38] and environment control (e.g., no oxygen) to NV magnetometry may allow measuring Cyt-C in different microbes such as *M. sedula*, *R. palustris*, and *M. acetivorans* under aerobic and anaerobic conditions.

Combining NV-$T_1$ magnetic relaxometry with NV-NMR[29–32] and NV-EPR[58] may enable analyzing Cyt-C inside cells (*in vivo*) or at least in the cell lysates (*in vitro*) by distinguishing the different spin signatures. Quantifying the changes in Cyt-C concentration within the cells is a crucial parameter for assessing cell damage. Consequently, observing the efflux of Cyt-C in living cells proves to be a convenient approach to address apoptosis initiation and monitoring the efficacy of strategies aimed at restoring apoptosis.[59] Using nanodiamonds with longer NV spin relaxation ($T_1$) and spin coherence ($T_2$) times[60] functionalized with cells may enable detection of intracellular Cyt-C concentrations.

**Associated content**

**Author Information**

**Author Contributions**

S.L. and R.T. performed NV measurements. C.S. synthesized the cytochrome C powder and functionalized the diamond surface. S.L. performed XPS, AFM, and SEM. I.F and A.L developed the NV-*T*1 relaxometry imaging technique. I.F. helped in analyzing and fitting the measured relaxometry images. K.A. performed EPR measurements. S.-H.L, R.Y.L., and A.L. conceived the cytochrome C study and supervised the project. All authors discussed the results. S.L. and A.L wrote the manuscript with contributions from all authors.

**Acknowledgement**

This material is based upon work supported by the NSF/EPSCoR RII Track-1: Emergent Quantum Materials and Technologies (EQUATE) Award OIA-2044049, and NSF Award 2328822. I.F. acknowledges support from Latvian Council of Science project lzp- 2021/1-0379. K.A. would like to acknowledge the support of the National Science Foundation/EPSCoR RII Track-4 Award OIA-2033210 and NSF Award 2328822. The research was performed in part in the Nebraska Nanoscale Facility: National Nanotechnology Coordinated Infrastructure and the Nebraska Center for Materials and Nanoscience (and/or NERCF), which are supported by NSF under Award ECCS: 2025298, and the Nebraska Research Initiative. We thank V.M. Acosta for suggesting the fabrication of non-metallic grating on diamond to prevent quenching effects caused by metals.

**Supporting Information**

**S1. Scanning electron microscopy analysis (SEM) of cytochrome C nanoclusters**

Figure S1a shows the SEM image performed on 2 µM cytochrome C (Cyt-C) drop-casted on the diamond substrate. To prevent charging issues, we sputtered the diamond substrate with Cyt-C with an ultrathin (thickness of 2.5 ± 0.5 nm) layer of Au. The diameter distribution of the Cyt-C nanoclusters is illustrated in the histogram depicted in Figure S1b indicating a mean diameter of ~ 100 ± 50 nm. The diameter is higher than the monomer (3.4 nm),[43] indicating that the aggregation is occurring.

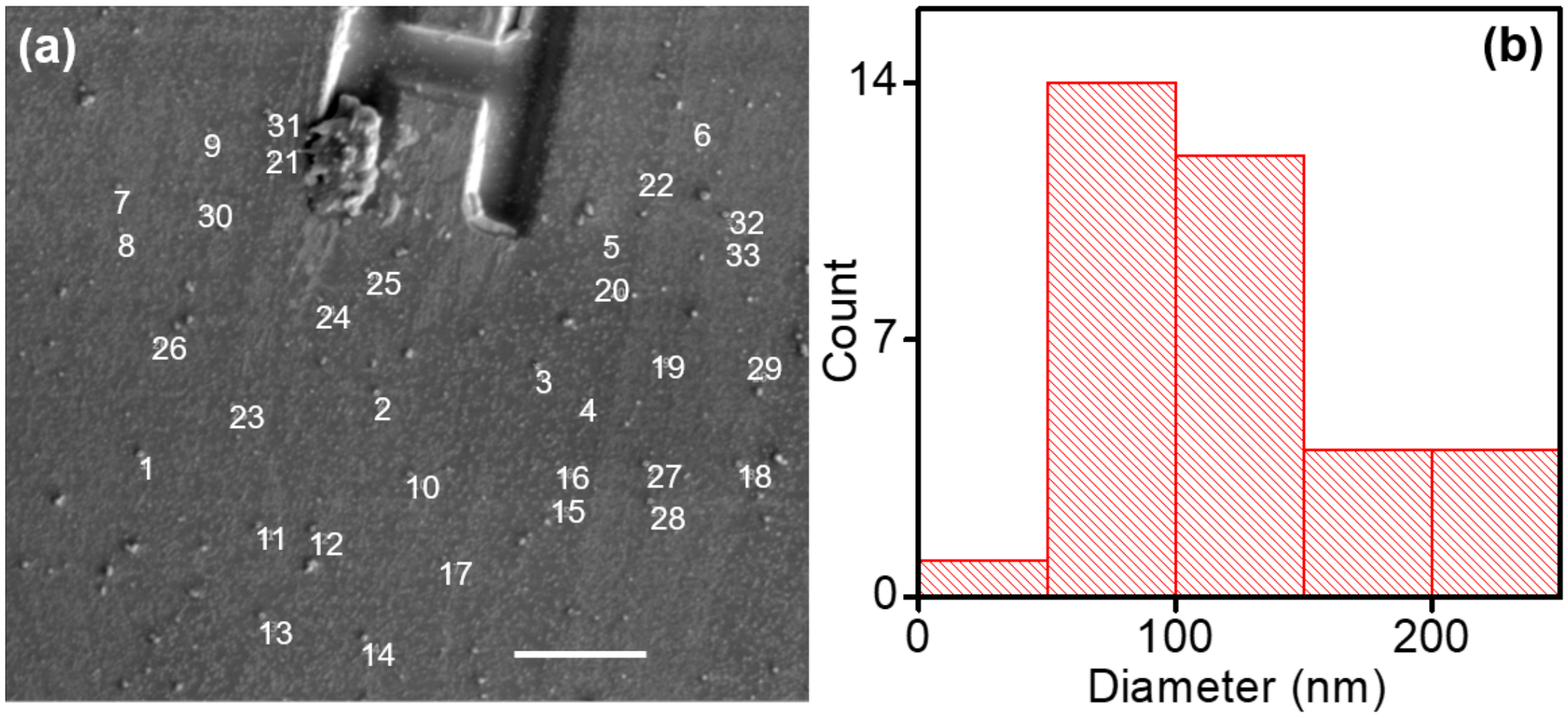

**Figure S1.** (a) SEM image of Cyt-C (concentration of 2 µM) drop-casted on the diamond chip and coated with ~ 2.5 ± 0.5 nm Au film to prevent charging effect. The scale bar in (a) is 5 µm. (b) The diameter distribution of Cyt-C nanoclusters with a diameter of 100 ± 50 nm.

**S2. X-ray photoelectron spectroscopy (XPS)**

We drop-casted 5 µL solution containing 54 µM Cyt-C onto a silicon substrate, allowing it to evaporate and form a thin film. Subsequently, X-ray photoelectron spectroscopy (XPS) analysis was conducted using Thermo Scientific Al K-Alpha XPS system, which operated under ultrahigh vacuum conditions at a pressure of $1 \times 10^{-9}$ mbar. Data acquisition and analysis were performed using the Avantage™ software package. To compensate for charge effects during measurement, a combination of electron and argon ion flood guns were employed, maintaining an argon pressure in the chamber ranging from $2 \times 10^{-8}$ to $4 \times 10^{-8}$ mbar. The X-ray beam had a diameter of 400 µm, and high-energy resolution spectra were recorded with a pass energy of 50 eV, utilizing a step size of 0.1 eV and a dwell time of 50 ms respectively. The number of averaged sweeps for each element was adjusted to optimize the signal to noise ratio, typically ranging from 20 to 100 sweeps.

For data analysis, we calibrated all spectra relative to a carbon 1s (Figure S2a) peak positioned at 284.8 eV to correct for any charging effects. Additionally, we performed background correction using the smart background subtraction feature within the Avantage™ package. Spectral analysis involved peak deconvolution, which utilized a combination of Gaussian and Lorentzian functions. The XPS results obtained for N 1s, O 1s, and S 2p are plotted in Figures S2b, S2c, and S2d. In the context of S 2p spectroscopy, we observed two distinct peaks. The first peak appears at 163.4 eV, indicating the presence of Fe-S bonding. This peak corresponds to the spectral line associated with the $2p_{1/2}$ electron configuration. The second peak occurring at 167 eV is associated with the spectral line of $2p_{3/2}$. This peak signifies the presence of sulfur bonded to the carbon-oxygen (C-O) chain.[61]

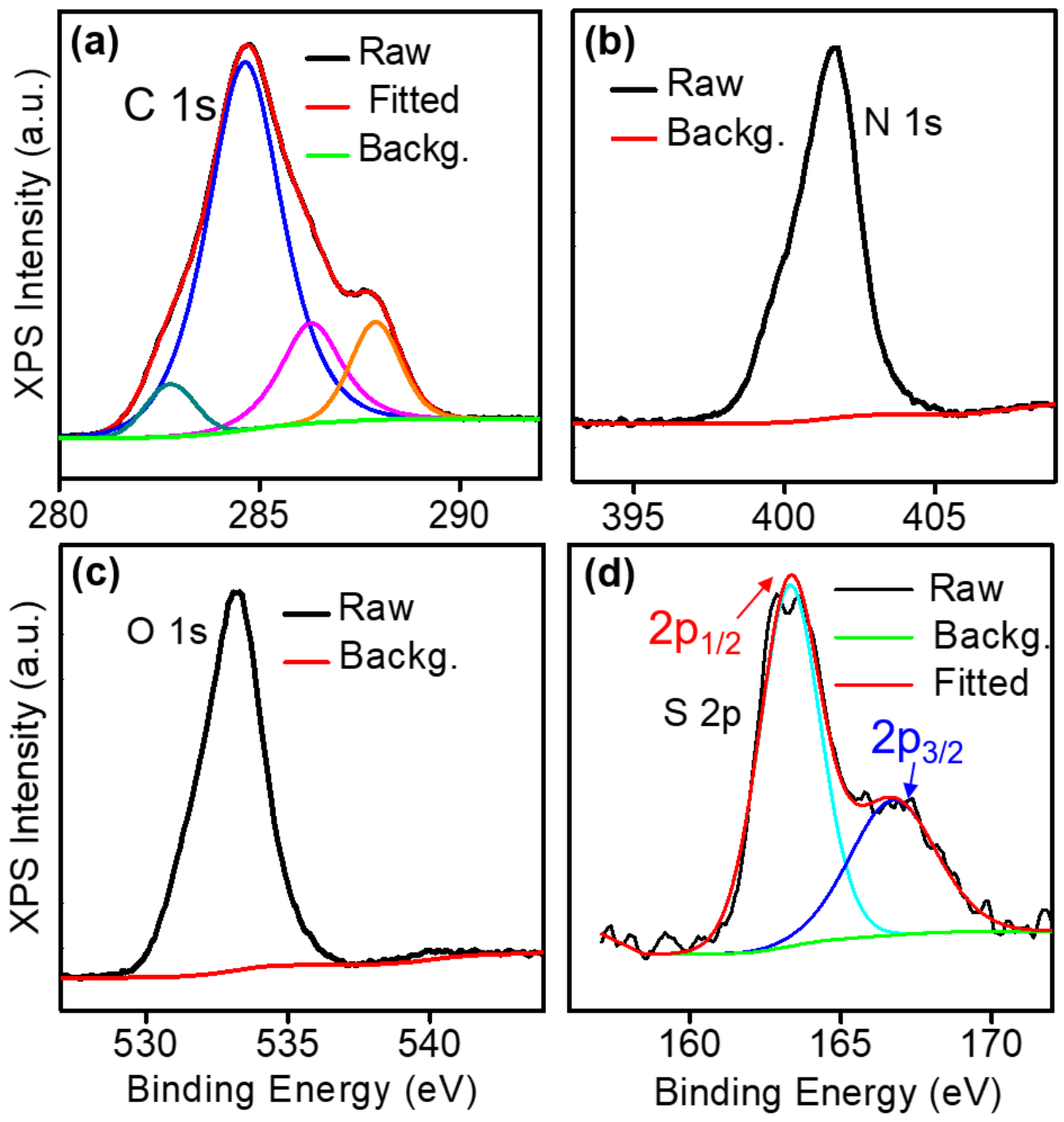

**Figure S2.** High resolution XPS peak with the background for (a) C 1s, (b) N 1s, (c) O1s, and (d) S 2p. The deconvoluted analysis for S 2p is also shown. There are two peak positions at 163 eV and 167 eV. 167 eV corresponds to the Fe-S binding energy.

**S3. Functionalization of the diamond substrate.**

To attach Cyt-C proteins, the NV-doped diamond chip was carboxylated at room temperature in a 9:1 mixture of concentrated $H_2SO_4$:$HNO_3$ for 72 hours. The diamond substrate was then rinsed and treated with 0.1 M HCl at 90 °C for 2 hours, followed by 0.1 M NaOH at 90 °C for 2 hours. The modification was confirmed using a Nicolet AVATAR 380 Fourier-transform infrared spectroscopy (FTIR) between 1000 $cm^{-1}$ to 3000 $cm^{-1}$. Peaks at 1960 $cm^{-1}$, 2030 $cm^{-1}$, 2160 $cm^{-1}$, and 2500 $cm^{-1}$ in the carboxylated diamond substrate are the intrinsic signature peaks for the diamond, Figure S3a. The two peaks at 2850 $cm^{-1}$ and 2920 $cm^{-1}$ showed an increase in hydrogen absorption on the surface. This facilitates the attachment of the Cyt-C molecules to the diamond surface. Figure S3b shows the $T_1$ relaxation measurement curves of the diamond before (filled circles) and after (open circles) the functionalization process. $T_1$ value is obtained as $1.24 \pm 0.1$ ms and $1.16 \pm 0.094$ ms before and after functionalization, respectively. The slight change in $T_1$ is related to the measurements done on a different region before and after carboxylation as there is a slight spatial change of the NV density, read the Supporting Information Section S5.

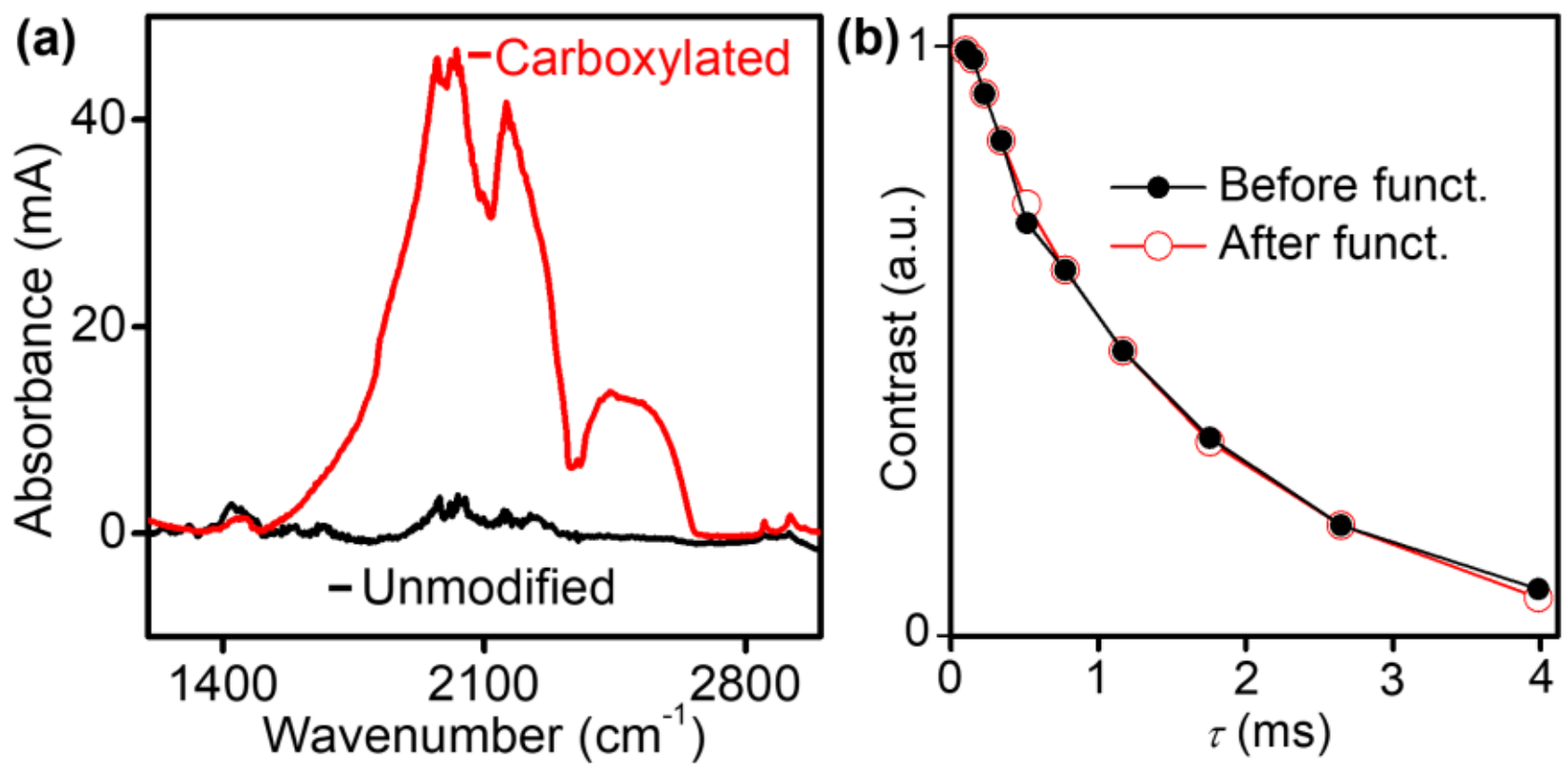

**Figure S3**. (a) FTIR spectroscopy of diamond chip before (red solid line) and after (black solid line) the functionalization of the diamond substrate. (b) $T_1$ relaxation of the diamond chip before (scattered filled-circle line) and after (scattered open-circle line) functionalization showing a slight change in relaxation times $T_1$ with the surface treatment.

## S4. Nanofabrication of SiN grating on top of diamond

We employed the lift-off technique to fabricate the grating structure on 4 mm × 4 mm × 0.5 mm diamond substrate. The diamond chip underwent a cleaning process involving a 200 °C treatment with tri-acid for a duration of 2 hours. Subsequently, a spin-coating process was utilized to apply a layer of LOR-3A, followed by the deposition of AZ-3312, which is a positive photoresist. To define the grating pattern, we utilized the Heidelberg DWL 66FS Laser Lithography system to expose the photoresist. After exposure, a development step was performed. To refine the surface, an ion beam milling process was employed to etch the 10 nm thick layer.

Following this, a 50 nm layer of silicon nitride (SiN) was sputter-deposited onto the substrate. The final step involved cleaning the diamond substrate to remove all the photoresist. The diamond with grating was subsequently cleaned in tri acid mixture at 200 °C. The dimension of SiN grid is 4 µm in width followed by a separation of 4 µm between the grids. Figure S4a shows the optical image of the grating on diamond with photoresist after DWL exposure and development. Figure S4b represents the AFM image of zoomed 20 µm by 20 µm grating, confirming the grating height of 50 ± 5 nm.

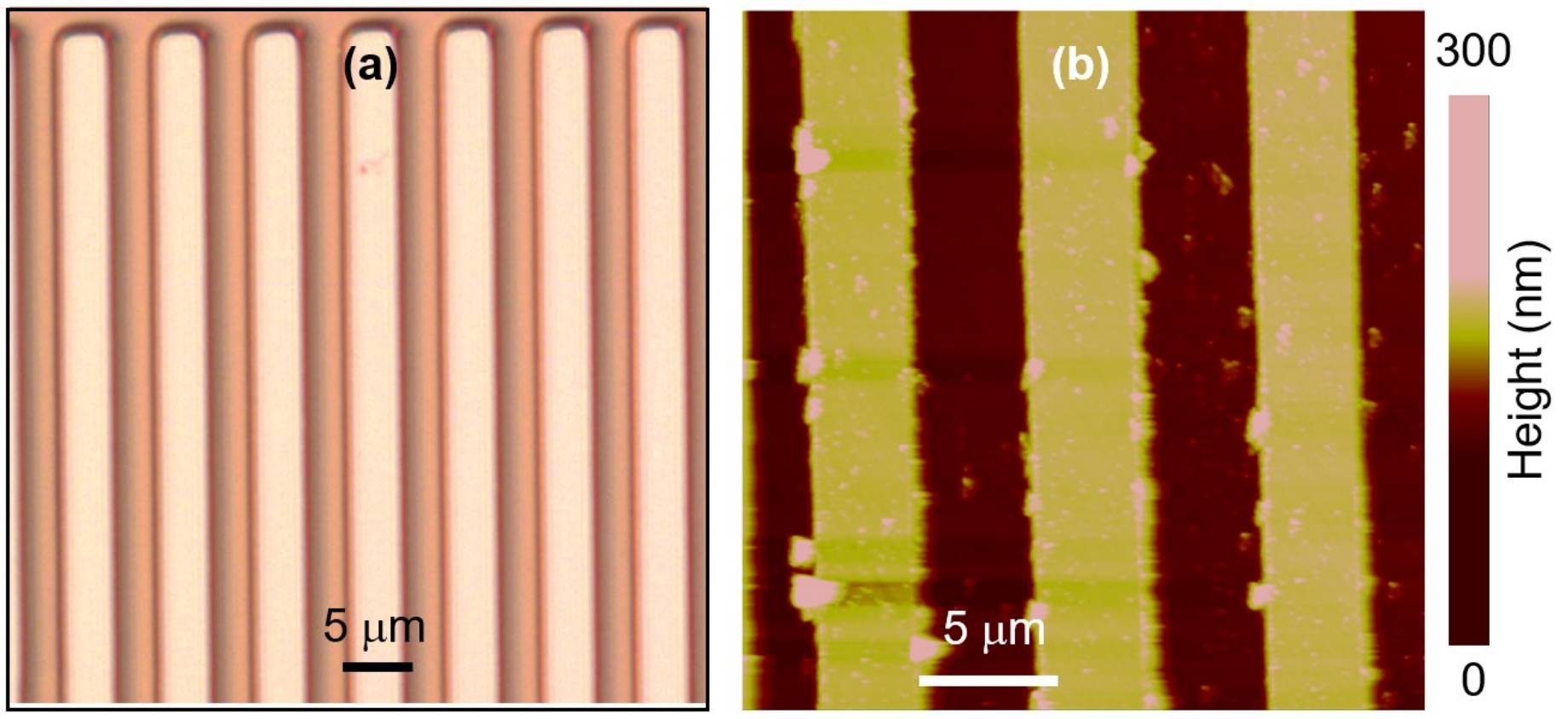

**Figure S4.** (a) Optical image of the SiN grating made on top of the diamond substrate after exposure with laser and development. (b) Zoomed AFM scan of the SiN grating with Cyt-C nanoclusters.

## S5. Experimental setup: optical detected magnetic resonance

***Creation of nitrogen vacancy (NV) centers in diamond.*** We used 4 mm × 4 mm × 0.5 mm type-IIa electronic grade (100) diamond (Element Six) substrate with a nitrogen concentration of < 5 ppb. We made first the SiN grating (Supporting Information Section S4) and then cut and polished the diamond substrate along the (100) plane at Delaware Diamond Knives, Inc. to obtain 4 mm × 2.5 mm × 0.05 mm membranes. The diamond is then implanted at CuttingEdge Ions with $^{15}N^{+}$ ions at an energy of 4 keV and a dose of $1 \times 10^{13}$ $cm^{-2}$ respectively to create a uniform layer of vacancies near the diamond surface ~ 5 – 8 nm.[18,50] We used Stopping and Range of Ions in Matter (SRIM) Monte Carlo simulations to estimate the vacancy distribution depth profile in the diamond and found a uniform distribution within ~ 8 nm beneath the diamond surface facing the $^{15}N^{+}$ source, Figure S5a. After the implantation we annealed the diamond substrate in an high vacuum (pressure ≤ $10^{-6}$ torr) furnace at 850 °C for 4 hours and at 1100 °C for 2 hours, and then cleaned it for two hours in a 1:1:1 mixture of nitric, sulfuric, and perchloric acid at 200 °C to remove graphite residues on the surface.[18,31,50] This process resulted in ~ 5 – 8 nm NV layer near the surface with a density of ~ 1 ppm, based on the fluorescence measurements.[62]

***Optical detected magnetic resonance setup.*** The wide field microscope (WFM) used in this study is a home-built optical detected magnetic resonance (ODMR) setup combined with sCMOS camera for wide-field imaging.[13,25] We used an oil immersion 100× Nikon objective with a working distance of 0.23 mm and NA of 1.25 to focus the pulsed green laser (532 nm) with a variable laser power (up to 2 W). The green laser pulsing is done by using Acousto-optical modulator (AOM) TEM-200-50-532 (Brimrose) and taking only the first order of diffraction laser mode. The carboxylated diamond substrate with 50-nm thick SiN grating and drop-casted Cyt-C is attached to a glass coverslip (thickness of 100 μm) with one microwave (MW) Au loop patterned on it for NV spin excitation. The coverslip with diamond is then mounted on a *x-y-z* motorized stage (Newport, Picometer 8742) that is connected to the computer through a USB interface for imaging a selected region of the sample. The fluorescence (650 – 800 nm) is collected by the same objective, transmitted through a dichroic mirror (DM, Semrock model #FF560-FDi01) and a single-band bandpass filter (Semrock FF01-731-137), and then focused on the sCMOS camera (Hamamatsu, ORCA-Flash4.0 V3) via a tube lens (Thorlabs, TTL200). For magnetic field alignment, the NV fluorescence is reflected by a flip mirror and focused with a lens (focal length of 30 mm) on an avalanche photon detector (APD, Thorlabs APD410A), connected to a Yokogawa oscilloscope (DL9041L). More details of the setup can be found in reference [13].

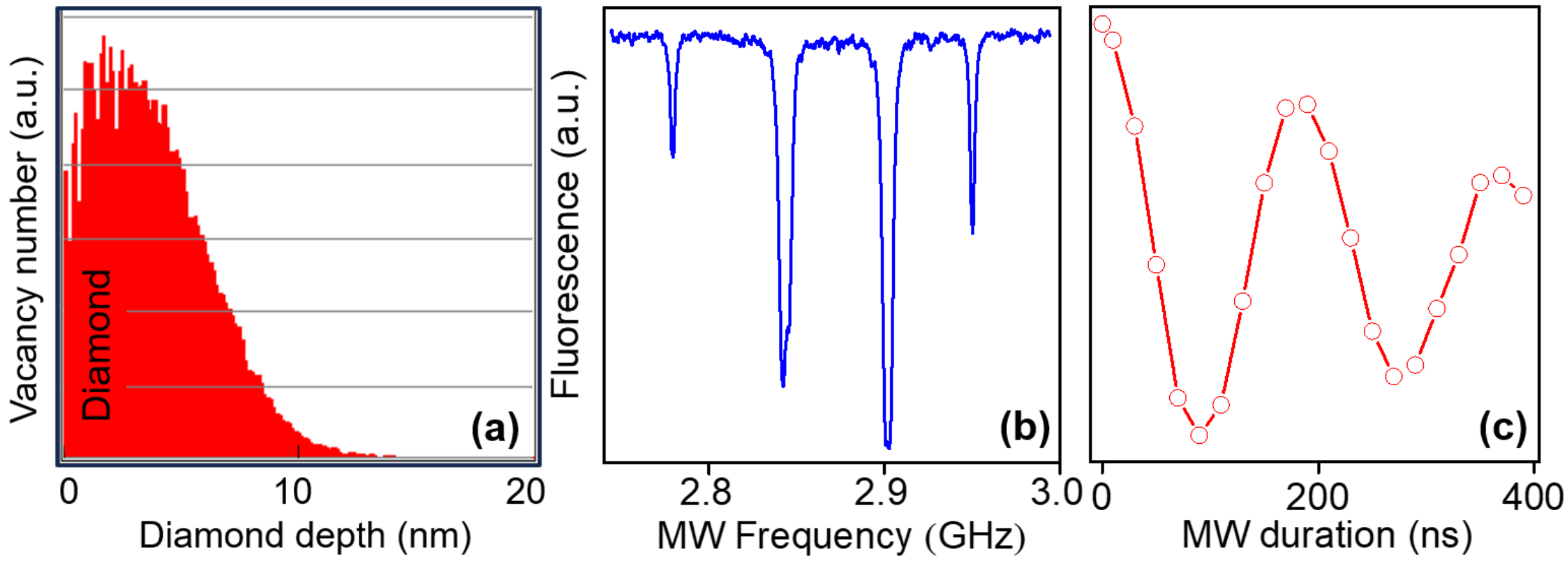

**Figure S5.** (a) Vacancy distribution obtained from SRIM calculations for 4 keV implanted $^{15}N$ in diamond (carbon) substrate. (b) ODMR spectrum of the diamond at $B_{app}$ of 3.2 mT. (c) Rabi oscillations of $m_s = 0$ to $m_s = -1$ spin transition at a MW frequency 2.78 GHz.

To perform pulsed (Rabi and $T_1$) NV measurements, we used a 400 MHz Pulse Blaster card (Spincore PBESR-PRO-400) to control the timing of green laser/MW pulses with the trigger of sCMOS camera for NV fluorescence detection/imaging. The MW pulses were generated by injecting MW AC current from a signal generator (SRS: Stanford Research Systems, model SG384) and directing it to a solid-state switch (Mini-Circuits ZASWA-2-50DR+), which was electrically operated by the TTL pulse provided by the pulse Blaster. Subsequently, these MW pulses were amplified using a Mini-Circuits ZHL-16W-43-s+ amplifier. We used permanent magnets (KJ magnetics, DX8X8-N52) to apply the magnetic field $B_{app}$ (up to 60 mT) along the NV [111] axis in the (100) diamond. By monitoring the ODMR peaks in the oscilloscope along [111] direction, we can align the applied magnetic field very accurately. **Figure S5b** shows four (two along [111] direction and two of the other merged NV peaks) ODMR peaks for $m_s = 0$ to $m_s = -1$ and $m_s = 0$ to $m_s = +1$ at $B_{app}$ of 3.2 mT. After the magnetic field alignment, we performed Rabi oscillations measurement of the $m_s = 0$ to $m_s = -1$ spins aligned along $B_{app}$ (3.2 mT) to obtain the $\pi$-pulse width, needed for $T_1$ measurement/imaging. The silver mirror was then flipped down to perform NV $T_1$ relaxometry imaging by the sCMOS camera, by varying the dark time $\tau$.[18] Read the main text for further details.

## S6. Data acquisition for $T_1$ relaxometry imaging

We used the following parameters for the camera: frame size is 550 pixels × 550 pixels (1 pixel = 65 nm at 100× magnification), the field of view is 36 µm × 36 µm. Contrast images were collected at various time $\tau$ points and averaged over many cycles (N = $10^4$) at each $\tau$ point. The single $\tau$ point images are shown in Figures S6b, S6c and S6d taken at $\tau$ of 50 µs, 300 µs and 1000 µs, respectively for a Cyt-C molecule of 18 µM drop-casted on the top of the diamond surface. The presence of $Fe^{+3}$ in Cyt-C generates spin noise, leading to a reduction in the $T_1$ relaxation time to 346 µs. Figure S3a depicts a decay curve, which was derived by assessing the average contrast from images captured at specific fixed $\tau$ values which was then fitted with an exponential decay curve keeping the amplitude and offset as free parameters to obtain the $T_1$ relaxation time.

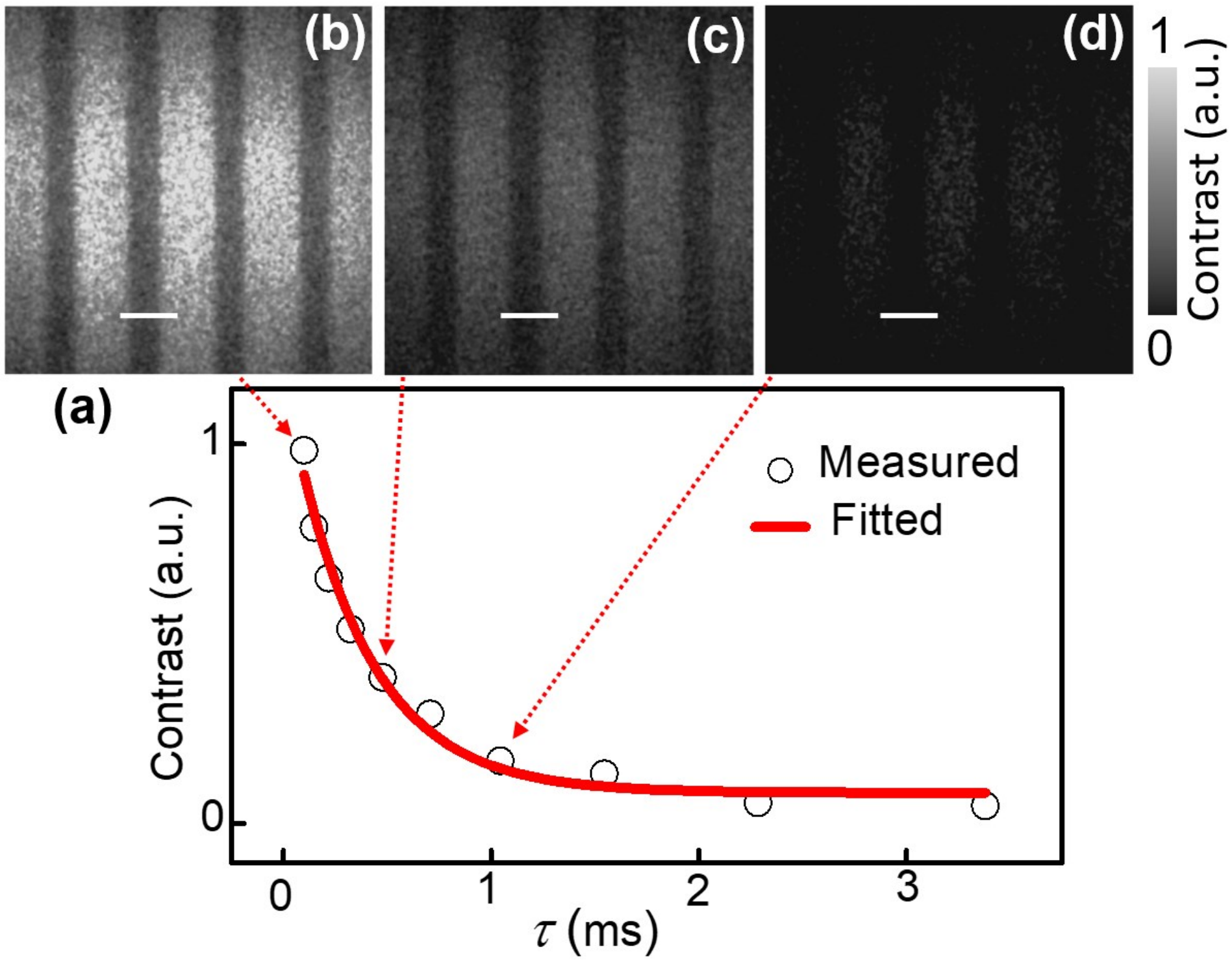

**Figure S6**. (a) Relaxation measurement (open circles) of NV spins in the presence of Cyt-C nanoclusters with a concentration of 18 µM, fitted (red solid line) with a decay exponential function. (b), (c), and (d) show the single point $\tau$ imaging at intervals of 50 µs, 300 µs, and 1000 µs, respectively. The scale bar in (b), (c), and (d) is 5 µm.

## S7. EPR data analysis.

Electron paramagnetic resonance (EPR) measurements were conducted using a Bruker EMX electron spin resonance setup equipped with an X-band metal cavity resonator (ER 4102ST cavity) at room temperature. The measurement conditions involved applying a 2 mT modulation magnetic field at a frequency of 100 kHz, while the MW frequency and power were set to 9.736 GHz and 20 mW respectively. EPR spectroscopy is employed to determine the *g*-factor of $Fe^{+3}$ present in Cyt-C proteins and $Fe^{+3}$ transverse spin relaxation $T_2^*$ rate ($\Gamma_2^* = 1/T_2^*$) by analyzing the linewidth of the EPR spectrum, as illustrated in Figure S7. By examining the first derivative of the EPR absorption spectrum of the pelletized powder Cyt-C (15 mg), we determined a full width at half maximum (FWHM) of 71.43 mT, corresponding to $\Gamma_2^* \sim 2$ GHz of the $Fe^{+3}$ spins.[34]

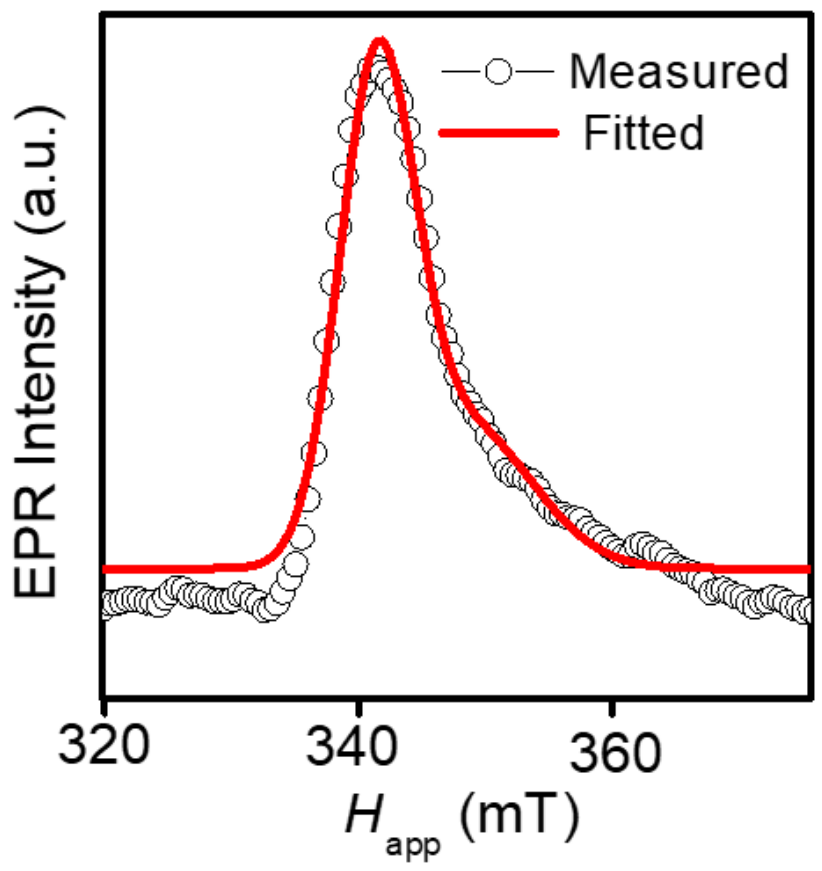

**Figure S7.** First derivative of the EPR absorption spectrum of the pelletized Cyt-C powder fitted with Gaussian functions to extract FWHM.

**S8. Theoretical estimation of relaxation rate of NV spins**

The heme center of Cyt-C produces a fluctuating magnetic field (or spin noise) that interacts with NV spins and reduces its $T_1$ relaxation time. The relaxation rate $\Gamma_1 = 1/T_1$ is known to be dependent upon *(i)* the dipolar interaction between external spins ($Fe^{+3}$ for our case) and NV spins and *(ii)* the fluctuation rate of the $Fe^{+3}$ spins in Cyt-C. This phenomenon is explained with the equation:[33]

$$\Gamma_1 = \frac{2\langle B^2\rangle f_t}{f_t + D^2}, \qquad \textbf{(Eq. S1)}$$

where, $f_t$ is the fluctuation rate of external spins, $\langle B^2\rangle$ is the mean dipolar magnetic coupling strength between the NV spins and the external $Fe^{+3}$ spins, and $D$ is zero field splitting of the NV spin ground state ~ 2.87 GHz. To calibrate our measurements and deduce the NV standoff $d$, we measured the relaxation rate $\Gamma_1$ in $CuSO_4$ solutions of different concentrations on the diamond substrate. We used equation Eq. S1 to estimate the theoretical relaxation rate of copper ions, Figure S8. The relaxation rate of the Cu ions is used as 0.3 GHz.[34] The dipolar magnetic field as a function of NV standoff *(d)* and concentration ($c$) is given as: [34]

$$10^3 \text{x} \frac{\pi c N_A}{8d^3} \text{ x} \left(\frac{\mu_0}{4\pi h} g_{NV}\, g_s\, \mu_B^2\right)^2, \qquad \textbf{(Eq. S2)}$$

where $g_{NV}$ and $g_S$ are the gyromagnetic ratio for NV spins and Cu ions respectively. $N_A$, $\mu_B$, and $h$ are Avogadro's number, Bohr magnetron, and Planck's constant respectively. $d$ is kept as a free parameter to match the fitting. As a result, we get $d$ of 5.5 nm.

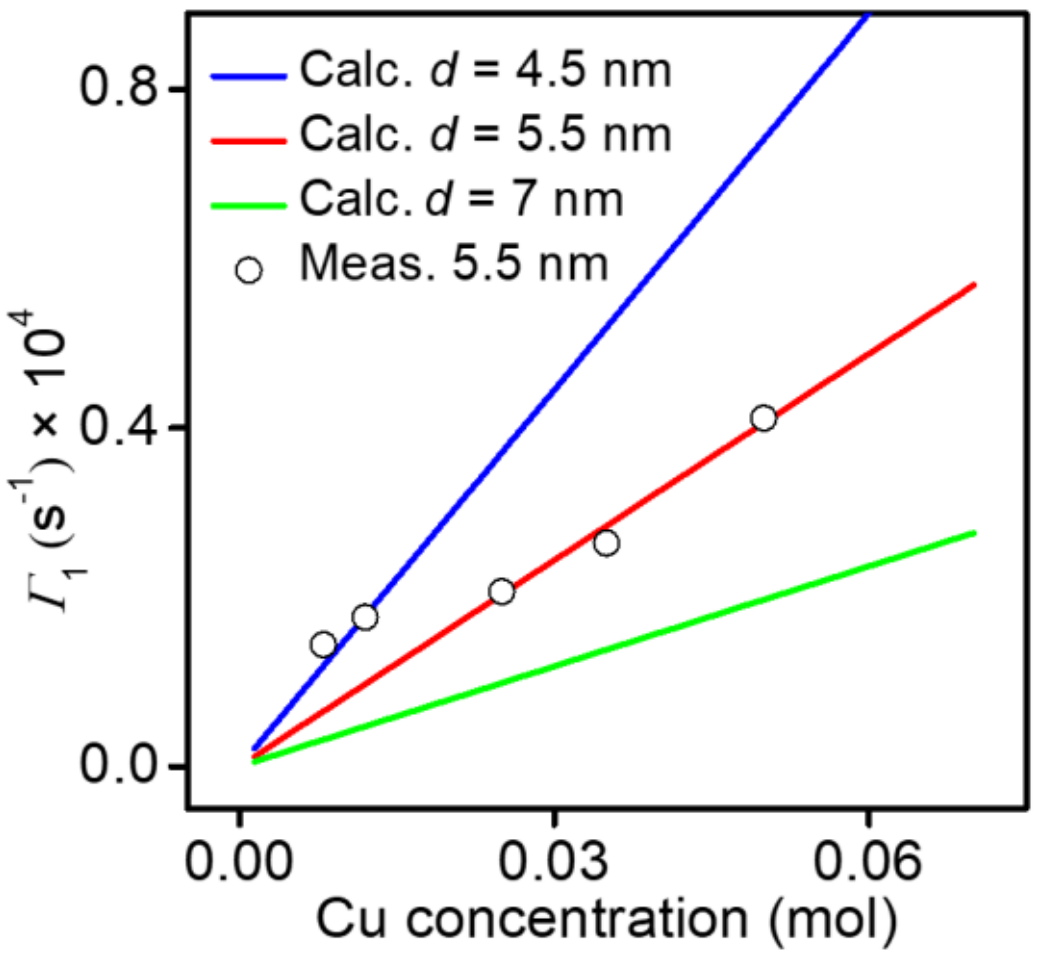

**Figure S8.** $\Gamma_1$ *vs* the concentration of Cu ions obtained both experimentally (open circles) and theoretically (solid lines: green: $d$ = 7 nm; red: $d$ = 5.5 nm; and blue: $d$ = 4.5 nm).

## References

(1) Bertini, I.; Cavallaro, G.; Rosato, A. Cytochrome *c* : Occurrence and Functions. *Chem. Rev.* **2006**, *106* (1), 90–115. https://doi.org/10.1021/cr050241v.

(2) Ow, Y.-L. P.; Green, D. R.; Hao, Z.; Mak, T. W. Cytochrome c: Functions beyond Respiration. *Nat Rev Mol Cell Biol* **2008**, *9* (7), 532–542. https://doi.org/10.1038/nrm2434.

(3) Liu, J.; Chakraborty, S.; Hosseinzadeh, P.; Yu, Y.; Tian, S.; Petrik, I.; Bhagi, A.; Lu, Y. Metalloproteins Containing Cytochrome, Iron–Sulfur, or Copper Redox Centers. *Chem. Rev.* **2014**, *114* (8), 4366–4469. https://doi.org/10.1021/cr400479b.

(4) Kühlbrandt, W. Structure and Function of Mitochondrial Membrane Protein Complexes. *BMC Biol* **2015**, *13* (1), 89. https://doi.org/10.1186/s12915-015-0201-x.

(5) Eleftheriadis, T.; Pissas, G.; Liakopoulos, V.; Stefanidis, I. Cytochrome c as a Potentially Clinical Useful Marker of Mitochondrial and Cellular Damage. *Front. Immunol.* **2016**, *7*. https://doi.org/10.3389/fimmu.2016.00279.

(6) Li, P.; Nijhawan, D.; Budihardjo, I.; Srinivasula, S. M.; Ahmad, M.; Alnemri, E. S.; Wang, X. Cytochrome c and dATP-Dependent Formation of Apaf-1/Caspase-9 Complex Initiates an Apoptotic Protease Cascade. *Cell* **1997**, *91* (4), 479–489. https://doi.org/10.1016/S0092-8674(00)80434-1.

(7) Radhakrishnan, J.; Wang, S.; Ayoub, I. M.; Kolarova, J. D.; Levine, R. F.; Gazmuri, R. J. Circulating Levels of Cytochrome *c* after Resuscitation from Cardiac Arrest: A Marker of Mitochondrial Injury and Predictor of Survival. *American Journal of Physiology-Heart and Circulatory Physiology* **2007**, *292* (2), H767–H775. https://doi.org/10.1152/ajpheart.00468.2006.

(8) Pyka, J.; Osyczka, A.; Turyna, B.; Blicharski, W.; Froncisz, W. EPR Studies of Iso-1-Cytochrome c : Effect of Temperature on Two-Component Spectra of Spin Label Attached to Cysteine at Positions 102 and 47. *European Biophysics Journal* **2001**, *30* (5), 367–373. https://doi.org/10.1007/s002490100168.

(9) Van Son, M.; Schilder, J. T.; Di Savino, A.; Blok, A.; Ubbink, M.; Huber, M. The Transient Complex of Cytochrome c and Cytochrome c Peroxidase: Insights into the Encounter Complex from Multifrequency EPR and NMR Spectroscopy. *ChemPhysChem* **2020**, *21* (10), 1060–1069. https://doi.org/10.1002/cphc.201901160.

(10) Blum, H.; Ohnishi, T. Electron Spin Relaxation of the Electron Paramagnetic Resonance Spectra of Cytochrome c. *Biochimica et Biophysica Acta (BBA) - Protein Structure* **1980**, *621* (1), 9–18. https://doi.org/10.1016/0005-2795(80)90057-4.

(11) Ichimura, K.; Nakahara, Y.; Kimura, K.; Inokuchi, H. Electronic States of Oxidized and Reduced Cytochrome c Observed by X-Ray Photoelectron Spectroscopy. *J. Mater. Chem.* **1992**, *2* (11), 1185. https://doi.org/10.1039/jm9920201185.

(12) Stavis, C.; Clare, T. L.; Butler, J. E.; Radadia, A. D.; Carr, R.; Zeng, H.; King, W. P.; Carlisle, J. A.; Aksimentiev, A.; Bashir, R.; Hamers, R. J. Surface Functionalization of Thin-Film Diamond for Highly Stable and Selective Biological Interfaces. *Proc. Natl. Acad. Sci. U.S.A.* **2011**, *108* (3), 983–988. https://doi.org/10.1073/pnas.1006660107.

(13) Lamichhane, S.; McElveen, K. A.; Erickson, A.; Fescenko, I.; Sun, S.; Timalsina, R.; Guo, Y.; Liou, S.-H.; Lai, R. Y.; Laraoui, A. Nitrogen-Vacancy Magnetometry of Individual Fe-Triazole Spin Crossover Nanorods. *ACS Nano* **2023**, *17* (9), 8694–8704. https://doi.org/10.1021/acsnano.3c01819.

(14) De Gille, R. W.; McCoey, J. M.; Hall, L. T.; Tetienne, J.-P.; Malkemper, E. P.; Keays, D. A.; Hollenberg, L. C. L.; Simpson, D. A. Quantum Magnetic Imaging of Iron Organelles within the Pigeon Cochlea. *Proc. Natl. Acad. Sci. U.S.A.* **2021**, *118* (47), e2112749118. https://doi.org/10.1073/pnas.2112749118.

(15) McCoey, J. M.; Matsuoka, M.; De Gille, R. W.; Hall, L. T.; Shaw, J. A.; Tetienne, J.; Kisailus, D.; Hollenberg, L. C. L.; Simpson, D. A. Quantum Magnetic Imaging of Iron Biomineralization in Teeth of the Chiton *Acanthopleura Hirtosa*. *Small Methods* **2020**, *4* (3), 1900754. https://doi.org/10.1002/smtd.201900754.

(16) Schäfer-Nolte, E.; Schlipf, L.; Ternes, M.; Reinhard, F.; Kern, K.; Wrachtrup, J. Tracking Temperature-Dependent Relaxation Times of Ferritin Nanomagnets with a Wideband Quantum Spectrometer. *Phys. Rev. Lett.* **2014**, *113* (21), 217204. https://doi.org/10.1103/PhysRevLett.113.217204.
(17) Wang, P.; Chen, S.; Guo, M.; Peng, S.; Wang, M.; Chen, M.; Ma, W.; Zhang, R.; Su, J.; Rong, X.; Shi, F.; Xu, T.; Du, J. Nanoscale Magnetic Imaging of Ferritins in a Single Cell. *Sci. Adv.* **2019**, *5* (4), eaau8038. https://doi.org/10.1126/sciadv.aau8038.
(18) Fescenko, I.; Laraoui, A.; Smits, J.; Mosavian, N.; Kehayias, P.; Seto, J.; Bougas, L.; Jarmola, A.; Acosta, V. M. Diamond Magnetic Microscopy of Malarial Hemozoin Nanocrystals. *Phys. Rev. Applied* **2019**, *11* (3), 034029. https://doi.org/10.1103/PhysRevApplied.11.034029.
(19) Doherty, M. W.; Manson, N. B.; Delaney, P.; Jelezko, F.; Wrachtrup, J.; Hollenberg, L. C. L. The Nitrogen-Vacancy Colour Centre in Diamond. *Physics Reports* **2013**, *528* (1), 1–45. https://doi.org/10.1016/j.physrep.2013.02.001.
(20) Laraoui, A.; Dolde, F.; Burk, C.; Reinhard, F.; Wrachtrup, J.; Meriles, C. A. High-Resolution Correlation Spectroscopy of 13C Spins near a Nitrogen-Vacancy Centre in Diamond. *Nat Commun* **2013**, *4* (1), 1651. https://doi.org/10.1038/ncomms2685.
(21) Laraoui, A.; Hodges, J. S.; Meriles, C. A. Magnetometry of Random Ac Magnetic Fields Using a Single Nitrogen-Vacancy Center. *Appl. Phys. Lett.* **2010**, *97* (14), 143104. https://doi.org/10.1063/1.3497004.
(22) Balasubramanian, G.; Neumann, P.; Twitchen, D.; Markham, M.; Kolesov, R.; Mizuochi, N.; Isoya, J.; Achard, J.; Beck, J.; Tissler, J.; Jacques, V.; Hemmer, P. R.; Jelezko, F.; Wrachtrup, J. Ultralong Spin Coherence Time in Isotopically Engineered Diamond. *Nature Mater* **2009**, *8* (5), 383–387. https://doi.org/10.1038/nmat2420.
(23) Degen, C. L.; Reinhard, F.; Cappellaro, P. Quantum Sensing. *Rev. Mod. Phys.* **2017**, *89* (3), 035002. https://doi.org/10.1103/RevModPhys.89.035002.
(24) Casola, F.; van der Sar, T.; Yacoby, A. Probing Condensed Matter Physics with Magnetometry Based on Nitrogen-Vacancy Centres in Diamond. *Nat Rev Mater* **2018**, *3* (1), 1–13. https://doi.org/10.1038/natrevmats.2017.88.
(25) Laraoui, A.; Ambal, K. Opportunities for Nitrogen-Vacancy-Assisted Magnetometry to Study Magnetism in 2D van Der Waals Magnets. *Appl. Phys. Lett.* **2022**, *121* (6), 060502. https://doi.org/10.1063/5.0091931.
(26) Erickson, A.; Shah, S. Q. A.; Mahmood, A.; Fescenko, I.; Timalsina, R.; Binek, C.; Laraoui, A. Nanoscale Imaging of Antiferromagnetic Domains in Epitaxial Films of Cr2O3 via Scanning Diamond Magnetic Probe Microscopy. *RSC Adv.* **2022**, *13* (1), 178–185. https://doi.org/10.1039/D2RA06440E.
(27) Aslam, N.; Zhou, H.; Urbach, E. K.; Turner, M. J.; Walsworth, R. L.; Lukin, M. D.; Park, H. Quantum Sensors for Biomedical Applications. *Nat Rev Phys* **2023**, 1–13. https://doi.org/10.1038/s42254-023-00558-3.
(28) Lovchinsky, I.; Sushkov, A. O.; Urbach, E.; de Leon, N. P.; Choi, S.; De Greve, K.; Evans, R.; Gertner, R.; Bersin, E.; Müller, C.; McGuinness, L.; Jelezko, F.; Walsworth, R. L.; Park, H.; Lukin, M. D. Nuclear Magnetic Resonance Detection and Spectroscopy of Single Proteins Using Quantum Logic. *Science* **2016**, *351* (6275), 836–841. https://doi.org/10.1126/science.aad8022.
(29) Degen, C. L.; Poggio, M.; Mamin, H. J.; Rettner, C. T.; Rugar, D. Nanoscale Magnetic Resonance Imaging. *Proc. Natl. Acad. Sci. U.S.A.* **2009**, *106* (5), 1313–1317. https://doi.org/10.1073/pnas.0812068106.
(30) Staudacher, T.; Shi, F.; Pezzagna, S.; Meijer, J.; Du, J.; Meriles, C. A.; Reinhard, F.; Wrachtrup, J. Nuclear Magnetic Resonance Spectroscopy on a (5-Nanometer) $^3$ Sample Volume. *Science* **2013**, *339* (6119), 561–563. https://doi.org/10.1126/science.1231675.
(31) Kehayias, P.; Jarmola, A.; Mosavian, N.; Fescenko, I.; Benito, F. M.; Laraoui, A.; Smits, J.; Bougas, L.; Budker, D.; Neumann, A.; Brueck, S. R. J.; Acosta, V. M. Solution Nuclear Magnetic Resonance Spectroscopy on a Nanostructured Diamond Chip. *Nat Commun* **2017**, *8* (1), 188. https://doi.org/10.1038/s41467-017-00266-4.

(32) Smits, J.; Damron, J. T.; Kehayias, P.; McDowell, A. F.; Mosavian, N.; Fescenko, I.; Ristoff, N.; Laraoui, A.; Jarmola, A.; Acosta, V. M. Two-Dimensional Nuclear Magnetic Resonance Spectroscopy with a Microfluidic Diamond Quantum Sensor. *Sci. Adv.* **2019**, *5* (7), eaaw7895. https://doi.org/10.1126/sciadv.aaw7895.
(33) Steinert, S.; Ziem, F.; Hall, L. T.; Zappe, A.; Schweikert, M.; Götz, N.; Aird, A.; Balasubramanian, G.; Hollenberg, L.; Wrachtrup, J. Magnetic Spin Imaging under Ambient Conditions with Sub-Cellular Resolution. *Nat Commun* **2013**, *4* (1), 1607. https://doi.org/10.1038/ncomms2588.
(34) Simpson, D. A.; Ryan, R. G.; Hall, L. T.; Panchenko, E.; Drew, S. C.; Petrou, S.; Donnelly, P. S.; Mulvaney, P.; Hollenberg, L. C. L. Electron Paramagnetic Resonance Microscopy Using Spins in Diamond under Ambient Conditions. *Nat Commun* **2017**, *8* (1), 458. https://doi.org/10.1038/s41467-017-00466-y.
(35) Nie, L.; Nusantara, A. C.; Damle, V. G.; Sharmin, R.; Evans, E. P. P.; Hemelaar, S. R.; Van Der Laan, K. J.; Li, R.; Perona Martinez, F. P.; Vedelaar, T.; Chipaux, M.; Schirhagl, R. Quantum Monitoring of Cellular Metabolic Activities in Single Mitochondria. *Sci. Adv.* **2021**, *7* (21), eabf0573. https://doi.org/10.1126/sciadv.abf0573.
(36) Wojtas, D.; Li, R.; Jarzębska, A.; Sułkowski, B.; Zehetbauer, M.; Schafler, E.; Wierzbanowski, K.; Mzyk, A.; Schirhagl, R. Quantum Sensing for Detection of Zinc-Triggered Free Radicals in Endothelial Cells. *Adv Quantum Tech* **2023**, *6* (11), 2300174. https://doi.org/10.1002/qute.202300174.
(37) Nie, L.; Nusantara, A. C.; Damle, V. G.; Baranov, M. V.; Chipaux, M.; Reyes-San-Martin, C.; Hamoh, T.; Epperla, C. P.; Guricova, M.; Cigler, P.; Van Den Bogaart, G.; Schirhagl, R. Quantum Sensing of Free Radicals in Primary Human Dendritic Cells. *Nano Lett.* **2022**, *22* (4), 1818–1825. https://doi.org/10.1021/acs.nanolett.1c03021.
(38) Wu, K.; Vedelaar, T. A.; Damle, V. G.; Morita, A.; Mougnaud, J.; Reyes San Martin, C.; Zhang, Y.; Van Der Pol, D. P. I.; Ende-Metselaar, H.; Rodenhuis-Zybert, I.; Schirhagl, R. Applying NV Center-Based Quantum Sensing to Study Intracellular Free Radical Response upon Viral Infections. *Redox Biology* **2022**, *52*, 102279. https://doi.org/10.1016/j.redox.2022.102279.
(39) Mugnol, K. C. U.; Ando, R. A.; Nagayasu, R. Y.; Faljoni-Alario, A.; Brochsztain, S.; Santos, P. S.; Nascimento, O. R.; Nantes, I. L. Spectroscopic, Structural, and Functional Characterization of the Alternative Low-Spin State of Horse Heart Cytochrome c. *Biophysical Journal* **2008**, *94* (10), 4066–4077. https://doi.org/10.1529/biophysj.107.116483.
(40) Miyashita, Y.; Wazawa, T.; Mogami, G.; Takahashi, S.; Sambongi, Y.; Suzuki, M. Hydration-State Change of Horse Heart Cytochrome c Corresponding to Trifluoroacetic-Acid-Induced Unfolding. *Biophysical Journal* **2013**, *104* (1), 163–172. https://doi.org/10.1016/j.bpj.2012.11.3825.
(41) Gong, Y.; Liu, Q.; Wilt, J. S.; Gong, M.; Ren, S.; Wu, J. Wrapping Cytochrome c around Single-Wall Carbon Nanotube: Engineered Nanohybrid Building Blocks for Infrared Detection at High Quantum Efficiency. *Sci Rep* **2015**, *5* (1), 11328. https://doi.org/10.1038/srep11328.
(42) Nucara, A.; Carbone, M.; Ripanti, F.; Manganiello, R.; Postorino, P.; Carbonaro, M. Achieving Cytochrome c Fibril/Aggregate Control towards Micro-Platelets and Micro-Fibers by Tuning pH and Protein Concentration: A Combined Morphological and Spectroscopic Analysis. *International Journal of Biological Macromolecules* **2019**, *138*, 106–115. https://doi.org/10.1016/j.ijbiomac.2019.07.060.
(43) Mirkin, N.; Jaconcic, J.; Stojanoff, V.; Moreno, A. High Resolution X-Ray Crystallographic Structure of Bovine Heart Cytochrome c and Its Application to the Design of an Electron Transfer Biosensor. *Proteins* **2007**, *70* (1), 83–92. https://doi.org/10.1002/prot.21452.
(44) Hirota, S.; Hattori, Y.; Nagao, S.; Taketa, M.; Komori, H.; Kamikubo, H.; Wang, Z.; Takahashi, I.; Negi, S.; Sugiura, Y.; Kataoka, M.; Higuchi, Y. Cytochrome *c* Polymerization by Successive Domain Swapping at the C-Terminal Helix. *Proc. Natl. Acad. Sci. U.S.A.* **2010**, *107* (29), 12854–12859. https://doi.org/10.1073/pnas.1001839107.
(45) Descostes, M.; Mercier, F.; Thromat, N.; Beaucaire, C.; Gautier-Soyer, M. Use of XPS in the Determination of Chemical Environment and Oxidation State of Iron and Sulfur Samples: Constitution of a Data Basis in Binding Energies for Fe and S Reference Compounds and Applications to the Evidence of Surface Species of an Oxidized Pyrite in a Carbonate Medium. *Applied Surface Science* **2000**, *165* (4), 288–302. https://doi.org/10.1016/S0169-4332(00)00443-8.
(46) Isaacson, Y. A.; Majuk, Z.; Brisk, M. A.; Gellender, M. E.; Baker, A. D. Iron to Sulfur Bonding in Cytochrome c Studied by X-Ray Photoelectron Spectroscopy. *J. Am. Chem. Soc.* **1975**, *97* (22), 6603–6604. https://doi.org/10.1021/ja00855a067.
(47) Jarmola, A.; Acosta, V. M.; Jensen, K.; Chemerisov, S.; Budker, D. Temperature- and Magnetic-Field-Dependent Longitudinal Spin Relaxation in Nitrogen-Vacancy Ensembles in Diamond. *Phys. Rev. Lett.* **2012**, *108* (19), 197601. https://doi.org/10.1103/PhysRevLett.108.197601.
(48) Mzyk, A.; Sigaeva, A.; Schirhagl, R. Relaxometry with Nitrogen Vacancy (NV) Centers in Diamond. *Acc. Chem. Res.* **2022**, *55* (24), 3572–3580. https://doi.org/10.1021/acs.accounts.2c00520.
(49) Ziem, F. C.; Götz, N. S.; Zappe, A.; Steinert, S.; Wrachtrup, J. Highly Sensitive Detection of Physiological Spins in a Microfluidic Device. *Nano Lett.* **2013**, *13* (9), 4093–4098. https://doi.org/10.1021/nl401522a.
(50) Kleinsasser, E. E.; Stanfield, M. M.; Banks, J. K. Q.; Zhu, Z.; Li, W.-D.; Acosta, V. M.; Watanabe, H.; Itoh, K. M.; Fu, K.-M. C. High Density Nitrogen-Vacancy Sensing Surface Created via He+ Ion Implantation of 12C Diamond. *Appl. Phys. Lett.* **2016**, *108* (20), 202401. https://doi.org/10.1063/1.4949357.
(51) Huang, L.-C. L.; Chang, H.-C. Adsorption and Immobilization of Cytochrome *c* on Nanodiamonds. *Langmuir* **2004**, *20* (14), 5879–5884. https://doi.org/10.1021/la0495736.
(52) Mrózek, M.; Rudnicki, D.; Kehayias, P.; Jarmola, A.; Budker, D.; Gawlik, W. Longitudinal Spin Relaxation in Nitrogen-Vacancy Ensembles in Diamond. *EPJ Quantum Technol.* **2015**, *2* (1), 1–11. https://doi.org/10.1140/epjqt/s40507-015-0035-z.
(53) Laraoui, A.; Hodges, J. S.; Meriles, C. A. Nitrogen-Vacancy-Assisted Magnetometry of Paramagnetic Centers in an Individual Diamond Nanocrystal. *Nano Lett.* **2012**, *12* (7), 3477–3482. https://doi.org/10.1021/nl300964g.
(54) Laraoui, A.; Meriles, C. A. Approach to Dark Spin Cooling in a Diamond Nanocrystal. *ACS Nano* **2013**, *7* (4), 3403–3410. https://doi.org/10.1021/nn400239n.
(55) Sangtawesin, S.; Dwyer, B. L.; Srinivasan, S.; Allred, J. J.; Rodgers, L. V. H.; De Greve, K.; Stacey, A.; Dontschuk, N.; O'Donnell, K. M.; Hu, D.; Evans, D. A.; Jaye, C.; Fischer, D. A.; Markham, M. L.; Twitchen, D. J.; Park, H.; Lukin, M. D.; de Leon, N. P. Origins of Diamond Surface Noise Probed by Correlating Single-Spin Measurements with Surface Spectroscopy. *Phys. Rev. X* **2019**, *9* (3), 031052. https://doi.org/10.1103/PhysRevX.9.031052.
(56) Ariyaratne, A.; Bluvstein, D.; Myers, B. A.; Jayich, A. C. B. Nanoscale Electrical Conductivity Imaging Using a Nitrogen-Vacancy Center in Diamond. *Nat Commun* **2018**, *9* (1), 2406. https://doi.org/10.1038/s41467-018-04798-1.
(57) Hall, L. T.; Kehayias, P.; Simpson, D. A.; Jarmola, A.; Stacey, A.; Budker, D.; Hollenberg, L. C. L. Detection of Nanoscale Electron Spin Resonance Spectra Demonstrated Using Nitrogen-Vacancy Centre Probes in Diamond. *Nat Commun* **2016**, *7* (1), 10211. https://doi.org/10.1038/ncomms10211.
(58) Shi, F.; Zhang, Q.; Wang, P.; Sun, H.; Wang, J.; Rong, X.; Chen, M.; Ju, C.; Reinhard, F.; Chen, H.; Wrachtrup, J.; Wang, J.; Du, J. Single-Protein Spin Resonance Spectroscopy under Ambient Conditions. *Science* **2015**, *347* (6226), 1135–1138. https://doi.org/10.1126/science.aaa2253.
(59) Pessoa, J. Live-Cell Visualization of Cytochrome *c* : A Tool to Explore Apoptosis. *Biochemical Society Transactions* **2021**, *49* (6), 2903–2915. https://doi.org/10.1042/BST20211028.
(60) Trusheim, M. E.; Li, L.; Laraoui, A.; Chen, E. H.; Bakhru, H.; Schröder, T.; Gaathon, O.; Meriles, C. A.; Englund, D. Scalable Fabrication of High Purity Diamond Nanocrystals with Long-Spin-Coherence Nitrogen Vacancy Centers. *Nano Lett.* **2014**, *14* (1), 32–36. https://doi.org/10.1021/nl402799u.
(61) B J Lindberg; K Hamrin; G Johansson; U Gelius; A Fahlman; C Nordling; K Siegbahn. Molecular Spectroscopy by Means of ESCA II. Sulfur Compounds. Correlation of Electron Binding Energy with Structure. *Phys. Scr.* **1970**, *1* (5–6), 286–298. https://doi.org/10.1088/0031-8949/1/5-6/020.

(62) Karki, P. B.; Timalsina, R.; Dowran, M.; Aregbesola, A. E.; Laraoui, A.; Ambal, K. An Efficient Method to Create High-Density Nitrogen-Vacancy Centers in CVD Diamond for Sensing Applications. *Diamond and Related Materials* **2023**, 110472. https://doi.org/10.1016/j.diamond.2023.110472.